\documentclass[aps,prx,reprint,a4paper,superscriptaddress,amsmath,amssymb,amsfonts,noshowpacs,longbibliography]{revtex4-2} 
\usepackage{newtxtext,newtxmath} 
\usepackage{physics}
\usepackage[T1]{fontenc}
\usepackage[utf8]{inputenx} 
\usepackage{graphicx} 
\usepackage[dvipsnames]{xcolor} 
\usepackage{soul} 
\usepackage[textwidth=17.5cm,textheight=23.5cm,verbose,pdftex]{geometry} 
\usepackage[pdftex]{hyperref}

\hypersetup{pdfauthor={Jonas Lähnemann}, bookmarksnumbered=true, pdftitle={Carrier diffusion in GaN---a cathodoluminescence study. III: Nature of nonradiative recombination at threading dislocations}, colorlinks, citecolor=blue, linkcolor=blue, urlcolor=blue}

\DeclareMathAlphabet{\mathcal}{OMS}{cmsy}{m}{n}

\begin{document} 
\title{Carrier diffusion in GaN---a cathodoluminescence study.\\ III: Nature of nonradiative recombination at threading dislocations}
\author{Jonas Lähnemann}
\email[Electronic mail: ]{laehnemann@pdi-berlin.de}
\author{Vladimir M. Kaganer}
\affiliation{Paul-Drude-Institut für Festkörperelektronik, Leibniz-Institut im Forschungsverbund Berlin e.\,V., Hausvogteiplatz 5--7, 10117 Berlin, Germany} 
\author{Karl K. Sabelfeld}
\author{Anastasya~E.~Kireeva}
\affiliation{Institute of Computational Mathematics and Mathematical Geophysics, Russian Academy of Sciences, Lavrentiev Prosp.~6, 630090 Novosibirsk, Russia} 
\author{Uwe Jahn}
\author{Caroline Chèze}
\author{Raffaella Calarco}
\altaffiliation{Present address: Istituto per la Microelettronica e Microsistemi, Consiglio Nazionale delle Ricerche, via del Fosso del Cavaliere 100, 00133~Roma, Italy}
\author{Oliver Brandt}
\email[Electronic mail: ]{brandt@pdi-berlin.de}
\affiliation{Paul-Drude-Institut für Festkörperelektronik, Leibniz-Institut im Forschungsverbund Berlin e.\,V., Hausvogteiplatz 5--7, 10117 Berlin, Germany}

\graphicspath{{./figs/}}

\begin{abstract}
We investigate the impact of threading dislocations with an edge component ($a$ or $a+c$-type) on carrier recombination and diffusion in GaN(0001) layers close to the surface as well as in the bulk. To this end, we utilize cathodoluminescence imaging of the top surface of a GaN(0001) layer with a deeply buried (In,Ga)N quantum well. Varying the acceleration voltage of the primary electrons and comparing the signal from the layer and the quantum well enables us to probe carrier recombination at depths ranging from the close vicinity of the surface to the position of the quantum well. Our experiments are accompanied by fully three-dimensional Monte Carlo simulations of carrier drift, diffusion, and recombination in the presence of the surface, the quantum well, and the dislocation, taking into account the dislocation strain field and the resulting piezoelectric field at the dislocation outcrop. Near the surface, this field establishes an exciton dead zone around the dislocation, the extent of which is not related to the carrier diffusion length. However, reliable values of the carrier diffusion length can be extracted from the dipole-like energy shift observed in hyperspectral cathodoluminescence maps recorded around the dislocation outcrop at low acceleration voltages. For high acceleration voltages, allowing us to probe a depth where carrier recombination is unaffected by surface effects, we observe a much stronger contrast than expected from the piezoelectric field alone. This finding provides unambiguous experimental evidence for the strong nonradiative activity of edge threading dislocations in bulk GaN and hence also in buried heterostructures.
\end{abstract}

\maketitle

\section{Introduction}
\label{sec:introduction} 

Much of the commercial success of the group-III nitrides for semiconductor devices is rooted in their relative tolerance against the presence of threading dislocations \cite{Rosner_1997, Hangleiter_2005, Chichibu_2006, Oliver_2010}. GaN-based light emitting diodes with external quantum efficiencies up to 80\% still have threading dislocation densities on the order of $10^8 - 10^9$~cm$^{-2}$ \cite{Nakamura_2015,Zhu_2016}, values that would be intolerable in other III-V semiconductors. Despite over two decades of research, the role of dislocations for the nonradiative recombination in GaN remains a controversial topic \cite{Jones_2000, Reshchikov_jap_2005, You_2009}.

In early studies, the comparatively benign nature of dislocations in GaN(0001) layers was attributed to the high ionicity of the material \cite{Lester_1995} or to the diffusion length being smaller than the average dislocation distance \cite{Rosner_1997, Speck_1999}. Alternatively, it was suggested that the dislocation spacing may actually be the decisive factor limiting the diffusion length \cite{Sugahara_1998, Bandic_2000, Chernyak_2001, Karpov_2003}. More recent works on free-standing GaN(0001) layers with very low dislocation densities, however, suggest that point defects are the limiting factor \cite{Scajev_2012,Sabelfeld_2017}. No consensus has been reached concerning the nonradiative nature of the different threading dislocations in GaN(0001). While some authors reported both $a$-type (edge) and $c$-type (screw) threading dislocations to be nonradiatively active, others reported either the former \cite{Hino_2000} or the latter \cite{Albrecht_2008} to be inactive. Several studies have also attributed specific  defect emission lines in GaN to the presence of dislocations \cite{Shreter_1997, Reshchikov_2005}, and thus associated them to radiative centers. 

At the first glance, for any dislocation, dangling bonds at the dislocation core are expected to create deep states in the gap that are likely to act as nonradiative centers. However, the core may reconstruct, removing dangling bonds and the associated electronic states. For each type of threading dislocation in GaN(0001), theoretical work has proposed a number of different core structures, some of which are expected to induce nonradiative centers, while others are not \cite{Arslan_2002, Belabbas_2007, Matsubara_2013}. Beyond the direct impact of the dislocation core, the long-range strain field of the dislocations may attract impurities and native point defects (Cottrell atmosphere \cite{Cottrell_1949}), which may possibly dominate the “outward” behavior of the dislocation \cite{Elsner_1998, Blumenau_2000, Lee_2000}. For example, defect-related luminescence associated with dislocations is commonly attributed to point defects in their vicinity \cite{Reshchikov_jap_2005}. In consequence, the specific effects of dislocations on the optical and electrical properties of a sample may depend on the growth method and growth conditions.

The strain field of edge threading dislocations \cite{Hirth_1982} induces a dipole-shaped shift of the band-gap energy around the dislocation line \cite{Gmeinwieser_2005, Gmeinwieser_2007,Liu_2016}. A screw component is accompanied by shear strain, but does not contribute to the normal strain components, and mixed dislocations thus exhibit essentially the same energy shift as edge threading dislocations. In piezoelectric materials such as GaN, the strain field of a dislocation generally induces a piezoelectric polarization, which in turn may give rise to electric fields. The threading dislocations relevant for GaN(0001) layers are actually an exception in this regard: for the wurtzite crystal structure, dislocations running parallel to the $\langle0001\rangle$ direction do not induce a piezoelectric field in the bulk, in contrast to dislocations along other directions \cite{Smirnova_1974,Shi_1999}. 

However, the sample surface constitutes a stress-free boundary, leading to a complex three-dimensional strain distribution also for $\langle0001\rangle$-oriented  dislocations \cite{Yoffe_1961}. This nonuniform strain field has been recognized only recently to result in a corresponding piezoelectric field around the outcrop of $a$-type edge dislocations \cite{Taupin_2014,Kaganer_2018}. This field may be strong enough to dissociate excitons and to spatially separate electrons and holes, thus inhibiting their radiative recombination up to a depth of about 140~nm \cite{Kaganer_2018}. Depending on the quadrant in which the exciton resides with respect to the dislocation at the moment of dissociation, the direction of the field is such as to repel electrons and to attract holes, or vice versa. Nonradiative recombination of electrons and holes can thus occur with electrons and holes being driven to the dislocation from different directions. As a consequence, for any experiment with a probe depth not significantly larger than these 140 nm, exciton field ionization will be the rate-limiting step instead of diffusion toward the dislocation line \cite{Kaganer_2019}.

Such experiments include the most popular ones for imaging dislocations, namely, cathodoluminescence (CL) and photoluminescence (PL) maps taken at the surface of GaN layers. The dark spots observed in these maps are invariably taken as direct evidence for the nonradiative activity of threading dislocations, including the edge (\emph{a}), mixed (\emph{a+c}) and screw (\emph{c}) types \cite{Rosner_1997, Sugahara_1998, Cherns_2001, Miyajima_2001, Liu_2016, Remmele_2001, Yamamoto_2003, Gmeinwieser_2005, Albrecht_2008, Naresh-Kumar_2014}. In contrast, for the edge and mixed types, the recent work of \citet{Taupin_2014} and \citet{Kaganer_2018,Kaganer_2019} suggests that the contrast observed stems from the piezoelectric field at the dislocation outcrop, and is only marginally affected by an additional nonradiative activity of the dislocation line \cite{Kaganer_2019}. However, it is the latter which is relevant for device structures used for light emitting diodes and diode lasers, where the active region is buried at a depth significantly exceeding the reach of the piezoelectric fields at the surface. Beyond the empirical knowledge that high densities of threading dislocations do indeed degrade the luminous efficacy of LEDs \cite{Nakamura_2015,Hurni_2015}, there is surprisingly little direct evidence for the nonradiative activity of threading dislocations in bulk GaN. 

The present work is the last part of a series of three papers; its two companion papers will be hereafter referred to as CD1 \cite{Jahn_2020} and CD2 \cite{Brandt_2020}. The investigated GaN/(In,Ga)N single quantum well (QW) sample is the same as in CD2, only that we are not performing CL measurements on the cross-section, but in top-view geometry, with the QW buried at a depth of about 650~nm. The Monte Carlo simulations accompanying these CL measurements build upon the companion paper CD1, in which the generation volume in GaN has been experimentally determined as a function of sample temperature and acceleration voltage. By varying the acceleration voltage and detection wavelength, we can probe the impact of threading dislocations on carrier recombination and diffusion from the surface to the well-defined depth of the QW, at which surface effects are almost negligible. For near-surface excitation, we determine the carrier diffusion length at temperatures between 10 and 300~K following the procedure introduced in Ref.~\citenum{Kaganer_2019}. The excellent agreement of the values with the values measured independently in CD2 \cite{Brandt_2020} on the same sample verifies that the understanding of the intensity contrast in terms of the piezoelectric field is sound. For increasing acceleration voltages, the intensity contrast remains higher than expected from the surface piezoelectric field alone. Our experiments thus firmly establish that threading dislocations in GaN with an $a$-type edge component act as nonradiative line defects in the bulk.

\section{Preliminary considerations} 
\label{sec:basics}

\begin{figure}
\includegraphics[width=0.85\columnwidth]{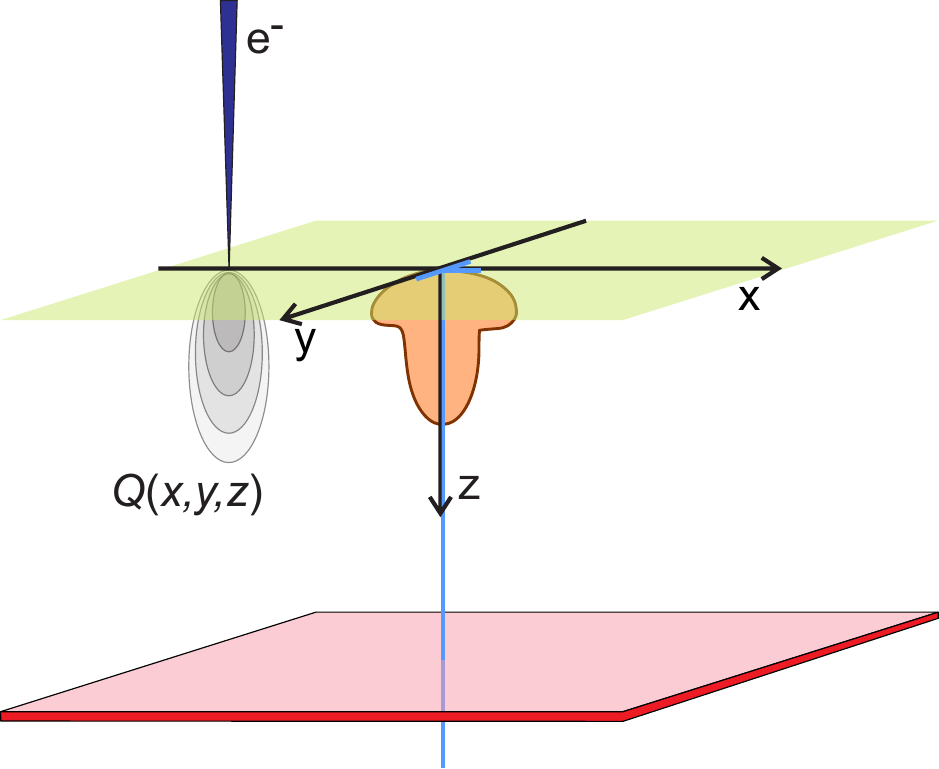}
	\caption{Configuration of the top-view CL experiment at the GaN(0001) surface (upper plane) with the electron beam scanning across the outcrop of a dislocation (vertical line), surrounded by a piezoelectric field (mushroom-like shape representing an isosurface of the field \cite{Kaganer_2018}) induced by the strain relaxation near the surface. The QW (bottom plane) is buried at a depth of 650~nm.}
\label{fig1} 
\end{figure}

In GaN, excitons and free carriers coexist at temperatures up to room temperature, with the ratio of their densities being controlled by the Saha equation \cite{Ebeling_1976}. In the present work, the term carriers will be used to subsume both excitons and free carriers. When we talk about the carrier diffusion length it thus represents a combination of exciton diffusion and ambipolar diffusion of electrons and holes, which we have investigated and discussed in detail in CD2 \cite{Brandt_2020} including the temperature dependence of the different contributions. Note that at room temperature, excitons in GaN constitute only a fraction of the total carrier population, and for the modest excitation density in the current experiments, this fraction amounts to only 10\%. However, from the lineshape and emission energy, we see that the experimentally recorded luminescence intensity is dominated by exciton recombination. This observation is understandable since the oscillator strength of the excitons is about 10$\times$ larger than for uncorrelated electron-hole pairs \cite{Brandt_1998}.

A schematic representation of our CL experiment is given in Fig.~\ref{fig1}. The sample is excited by the focused electron beam impinging on the top surface along the $z$ direction. The carrier generation volume $Q(x,y,z)$ (source) is the region where electron-hole pairs are excited by the scattered electrons. Its dimensions depend on the acceleration voltage $V$ and sample temperature $T$ \cite{Jahn_2020}. The beam is scanned across the surface along the $x$ and $y$ directions around the dislocation outcrop at $x=y=0$. Either hyperspectral maps of the GaN near-band-edge emission or monochromatic maps of the emission intensity at a fixed wavelength are recorded. Close to the dislocation, excitons dissociate and carriers are separated in the piezoelectric field resulting from the relaxation of the dislocation strain near the surface, which inhibits their radiative recombination \cite{Kaganer_2018}. The same strain field produces a band gap variation that is reflected in the emission energy around the dislocations \cite{Gmeinwieser_2005,Liu_2016,Kaganer_2019}. On the tensile side, the emission will be redshifted, while on the compressive side, the emission will be blueshifted. Taking the dissociation of excitons in the piezoelectric field into account, the effective lifetime of the excitons is given by
\begin{equation}
\frac{1}{\tau_\mathrm{eff}} = \frac{1}{\tau_\mathrm{r}} + \frac{1}{\tau_\mathrm{nr}} + \frac{1}{\tau_E[E(x,y,z)]},
\end{equation}
where $\tau_\mathrm{r}$ and $\tau_\mathrm{nr}$ are the radiative and nonradiative lifetimes far from the dislocation, respectively,  while, according to Eq.~(8) in Ref.~\citenum{Kaganer_2018}, $1/\tau_E$ is the position-dependent dissociation rate of the free exciton in the electric field $\vb*{E}(x,y,z)$ with the magnitude $E = \abs{\vb*{E}}$ calculated in analogy to the field ionization probability (per unit time) of the hydrogen atom. The effective carrier lifetime in our sample, which is essentially the bulk nonradiative lifetime far from the dislocations, was measured in CD2 \cite{Brandt_2020} to range from 15~ps at 300~K to 80~ps at 10~K. $\tau_E$ is smaller than these values for fields above 12.2 and 10.4~kV/cm, respectively. According to Ref.~\citenum{Kaganer_2018}, an isosurface for $E=10$~kV/cm has a radius of around 70~nm close to the surface and extends to a depth of about 140~nm. For $V \lessapprox 5$~kV, the generation volume does not extend beyond the range of this field.

\begin{figure} 
\includegraphics[width=1\columnwidth]{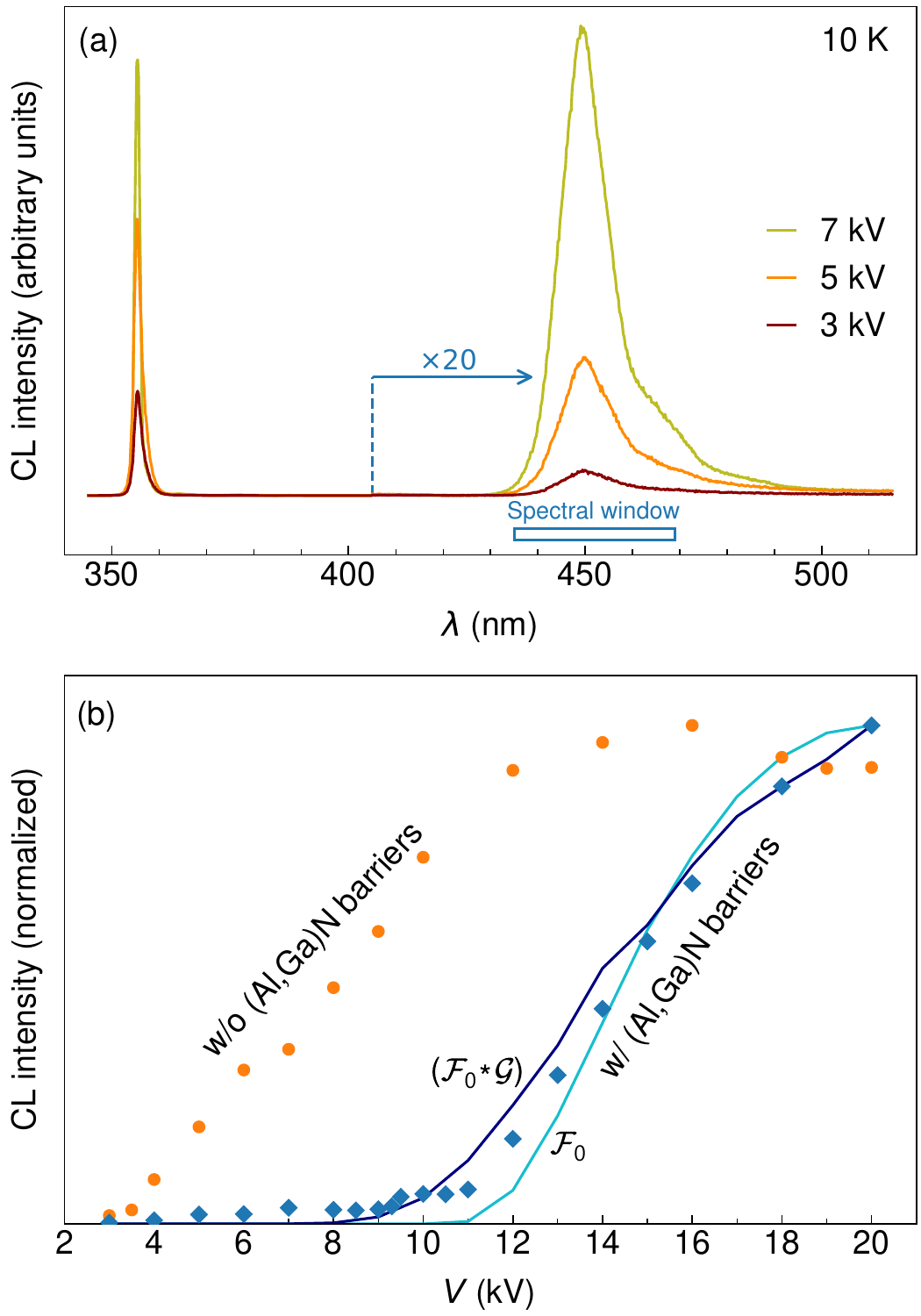} 
	\caption{(a) CL spectra acquired while scanning an area of 97~\textmu{}m$^2$ on the surface of the sample under investigation for different acceleration voltages $V$ at 10~K. The spectra are normalized by the respective electron beam currents. For the QW emission, the spectral window used for recording monochromatic CL photon counting maps is indicated. (b) Integrated QW emission intensity as a function of acceleration voltage $V$ for the sample under investigation (circles) compared to a sample where diffusion to the QW is inhibited by additional (Al,Ga)N barriers (diamonds). The data are normalized to the respective electron beam currents to match the situation of a constant number of electrons used in the simulations \cite{Toth_1998}. The intensity of the sample with additional barriers is compared to two calculated dependencies having no free parameter. First, the curve labeled $\mathcal{F}_0$ is calculated from the energy loss depth profile obtained by simulations using the software \texttt{CASINO}. Second, for the curve labeled $\mathcal{F}_0 * \mathcal{G}$, we take into account a voltage-dependent increase of the CL generation volume by a convolution with a Gaussian $\mathcal{G}$.}
	\label{figD}
\end{figure}

It is crucial for our investigation that the QW is not directly excited by the electron beam but only by carrier diffusion from the generation volume to the QW. Only in this case, the spatial distribution of the QW emission serves as a probe for the carrier distribution in the GaN layer above and gives insights into the activity of the dislocation line in bulk GaN. Low-temperature CL spectra recorded for different acceleration voltages $V$ are displayed in Fig.~\ref{figD}(a). The spectra show two CL lines centered at 357 and 450~nm, which can be attributed to the GaN matrix and the In$_{0.16}$Ga$_{0.84}$N QW, respectively. To account for any inhomogeneities of the sample, monochromatic maps used to extract intensity profiles are recorded with a large spectral bandwith of 33~nm as indicated for the QW emission in Fig.~\ref{figD}(a).

It is clear from Fig.~\ref{figD}(a) that with increasing $V$ and thus larger penetration depth, the contribution from the QW emission increases strongly. Figure~\ref{figD}(b) shows that the integrated intensity from the QW emission band increases essentially linearly from 3 to 12~kV, and saturates for higher values of $V$. For comparison, the same measurement has been repeated for the sample from CD1 \cite{Jahn_2020}, where the QW is sandwiched between additional Al$_{0.11}$Ga$_{0.89}$N barriers that inhibit carrier diffusion to the QW. This second curve is thus governed by direct excitation of the QW or carriers excited in the barriers that reach the QW, which evidently sets in for $V>10$~kV only (the finite signal for lower values of $V$ originates from the optical excitation of the QW by photons originating from the GaN cap layer). Clearly, the signal of the QW in the sample without additional barriers is induced by carrier diffusion for all but the highest values of $V$.

\begin{figure*}
\includegraphics[width=1\textwidth]{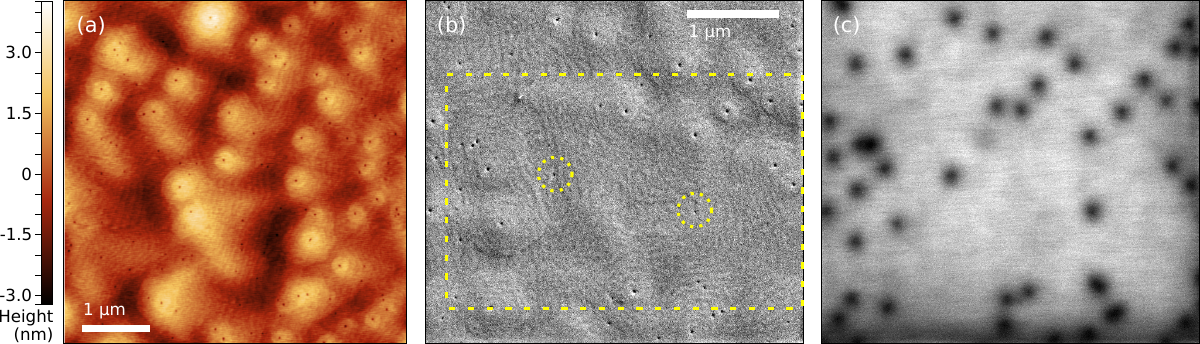}
	\caption{(a) Atomic force topograph of the surface of the investigated sample. (b) Top-view secondary electron micrograph of the sample surface and (c) corresponding monochromatic CL map of the near-band-edge emission. The dashed rectangle in (b) marks the area on which the hyperspectral map presented in Fig.~\ref{figMap} was recorded and the dotted circles highlight the investigated dislocations.}
\label{figExp} 
\end{figure*}

It is interesting to have a closer look at the signal of the sample with barriers, since it contains information on the vertical extent of the generation volume. In fact, the signal can be modeled based on the energy loss profile in $z$ direction computed by \texttt{CASINO} \cite{Demers_2011} with the same parameters as used in CD1 \cite{Jahn_2020}. Integrating these profiles over the QW and (Al,Ga)N barriers---a region of 33~nm centered at a depth of 650~nm---we obtain the curve labeled $\mathcal{F}_0$ as shown in Fig.~\ref{figD}(b). 
The onset of $\mathcal{F}_0$ occurs at slightly higher $V$ than in the experiment, while the slope is steeper. As we have shown in CD1 \cite{Jahn_2020}, the lateral extent of the generation volume for CL is underestimated by \texttt{CASINO}. The same should apply to the vertical extent of the generation volume. In fact, convoluting $\mathcal{F}_0$ with a Gaussian $\mathcal{G}$ prior to the integration over the QW and barriers comes even closer to the experimental data (note that the calculation has no free parameter). Since the width $\sigma$ of the Gaussian describes the quasi-diffusive spread of the carrier distribution during the cooldown of hot carriers to the band edges, it naturally should be similar in all directions. We thus take this phenomenological parameter from Fig.~10(b) in CD1 \cite{Jahn_2020} and extrapolate it to higher $V$ assuming that it follows a linear dependence on $V$.

\section{Experiment} 
\label{sec:experiment}

For the present experiments, we utilize a 1.3-\textmu m-thick GaN layer fabricated by plasma-assisted molecular beam epitaxy (MBE) on top of a GaN(0001) template, which in turn was prepared by metal-organic chemical vapor deposition on an Al$_2$O$_3$(0001) substrate. A 3-nm-thick In$_{0.16}$Ga$_{0.84}$N single QW is buried at about $650$~nm below the surface. This sample is the same as the one investigated in the preceeding study (CD2 \cite{Brandt_2020}), but we now collect CL from the top surface instead of from the cross-section of the sample. Additional measurements for Fig.~\ref{figD}(b) were carried out on a companion sample, grown under nominally identical conditions, where the QW is sandwiched between two additional 15~nm-wide Al$_{0.11}$Ga$_{0.89}$N barriers (the same sample as studied in CD1 \cite{Jahn_2020}).

The CL experiments were performed using a Gatan mono\-CL4 system and a He-cooling stage attached to a Zeiss Ultra55 scanning electron microscope with a field emission gun. Unless otherwise noted, the acceleration voltage $V$ of the incident primary electrons and the beam current were set to 3~kV (penetration depth of $\approx 50$~nm) and 0.85~nA, respectively. The low value of $V$ was chosen to achieve a high spatial resolution. Hyperspectral maps of the GaN emission around the outcrops of disclocations were recorded using a charge coupled device detector. A grating with 1200~lines/mm and input slit widths of 0.1, 0.3, and 0.5~mm, corresponding to spectral resolutions of 0.3, 0.9 and 1.5~nm, were chosen for temperatures of 10, 60 and 120~K and above, respectively. The hyperspectral images were analyzed using the \texttt{PYTHON} package \texttt{HyperSpy} \cite{Hyperspy}. Additional CL intensity profiles across a dislocation core were obtained from monochromatic CL photon counting maps acquired for temperatures $T$ between 10 and 300~K, varying $V$ between 3 and 10~kV. Here, a grating with 300~lines/mm was used and the slits were opened to 3~mm to obtain a wide spectral window of 33~nm covering either the near-band-edge emission of GaN or the broader emission band of the buried (In,Ga)N quantum well. The beam current was reduced at both lower $T$ and higher $V$ and neutral density filters with optical densities between 0.3 and 2 were used to attenuate the signal to the optimum operating range of the photomultiplier tube. CL intensity profiles extracted from these maps were integrated over 60~nm wide stripes. To compare the intensity contrast at the dislocation, all intensity profiles presented in this study are normalized to 1 far from the dislocation. Overview spectra at low temperature were obtained using the same grating, but with a reduced slit width of 0.1~mm ($\approx 1.2$~nm spectral resolution).

Simulations of both the CL emission intensity and energy are carried out using an advanced Monte Carlo approach \cite{Sabelfeld_1991}. The solution for the pure diffusion problem \cite{Sabelfeld_2016} has been extended to include the drift of excitons in the strain field of an $a$-type threading dislocation and to account for the dissociation of excitons in the resulting piezoelectric field \cite{Kaganer_2019}. The full three-dimensional drift and diffusion of excitons is simulated. Details on the algorithm can be found in the Appendix of Ref.~\citenum{Kaganer_2019}. The value of the shear piezoelectric coefficient is taken as $e_{15} = - e_{31}$ \cite{Kaganer_2018, Kaganer_2019}, and the surface-recombination velocity is set to $S = 500$~nm/ns \cite{Aleksiejunas_2003}. The diffusion length $L=\sqrt{D\tau}$ is the only parameter that is adjusted in our calculations. As the carrier lifetime far from the dislocation is known from CD2 \cite{Brandt_2020}, we essentially vary the diffusivity $D$. To study the influence of the dislocation in the bulk, it is modeled as a cylinder of radius $R=3$~nm with an infinite recombination velocity on the cylinder surface. Such a treatment was first proposed in the 1970s \cite{vanOpdorp_1977, Lax_1978, Donolato_1998} and has since been repeatedly employed also for GaN \cite{Yakimov_2002, Sabelfeld_2017}. The choice of the dislocation radius is motivated by the exciton Bohr radius in GaN. 

Figure~\ref{figExp}(a) shows an atomic force topograph of the surface of our sample. The atomic steps on the surface are clearly discernible, and  dislocation outcrops are visible as nm-sized pits. A similar view of the surface can be obtained from secondary electron micrographs such as the one shown in Fig.~\ref{figExp}(b), where again both atomic steps on the surface and the outcrops of dislocations can be recognized easily. Figure~\ref{figExp}(c) presents a monochromatic CL map of the near-band-edge emission of GaN at 10~K recorded simultaneously with the image in Fig.~\ref{figExp}(b). All the dislocations visible in Fig.~\ref{figExp}(b) show up in the CL map as extended dark spots with a diameter much larger than the actual pit size seen in the atomic force and secondary electron micrographs. The spatial extent of the CL contrast is thus not determined by the pit size, as it might be the case for samples exhibiting the characteristic $\vee$ pits often found for (In,Ga)N quantum wells \cite{Sharma_2000}.

The average dislocation density determined from a number of such atomic force topographs and CL maps amounts to $5\times 10^8$~cm$^{-2}$, which is similar to typical densities in structures used for commercial light emitting diodes and other GaN-based devices \cite{Zhu_2016}. Note that the two isolated threading dislocations in the central part of Fig.~\ref{figExp}(b) are likely pure edge dislocations, as they do not sit on the apex of a hillock, but between the monoatomic steps on the sample surface \cite{Han_2008}. Dislocations with a spiral hillock around their outcrop clearly contain a screw component, but we cannot distinguish between either mixed or pure screw dislocations.

\section{Results and Discussion}
\label{sec:results}

\subsection{Hyperspectral CL maps around dislocations}
\label{sec:hyperspec}

\begin{figure} 
\centering{\includegraphics[width=1\columnwidth]{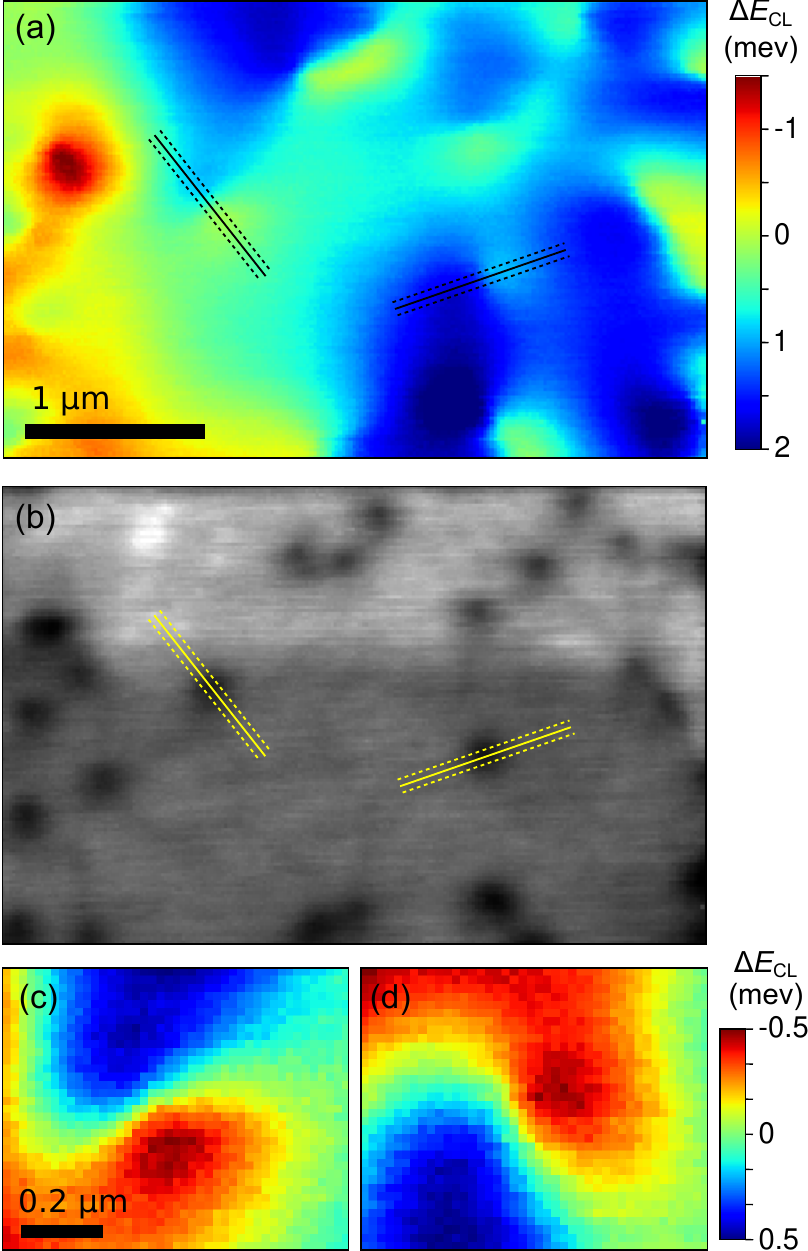}} 
\caption{(a) Energy shift $\Delta E_\mathrm{CL}$ and (b) relative CL intensity obtained from a Lorentzian fit to the near-band-edge line in a hyperspectral CL map recorded on the sample surface at $T=10$~K and $V=3$~kV. The dark spots in (b) reveal the positions of single dislocations. Each of these dislocations is associated with a dipole-like shift in emission energy in (a), further highlighted in (c) and (d) for the two isolated dislocations in the center of the map. Profiles of the emission energy and intensity for these two dislocations are extracted along the marked lines and presented in Fig.~\ref{figE}, with the dashed lines identifying the width over which the signal was integrated.}
\label{figMap}
\end{figure}

\begin{figure*}
\centering{\includegraphics[width=1\textwidth]{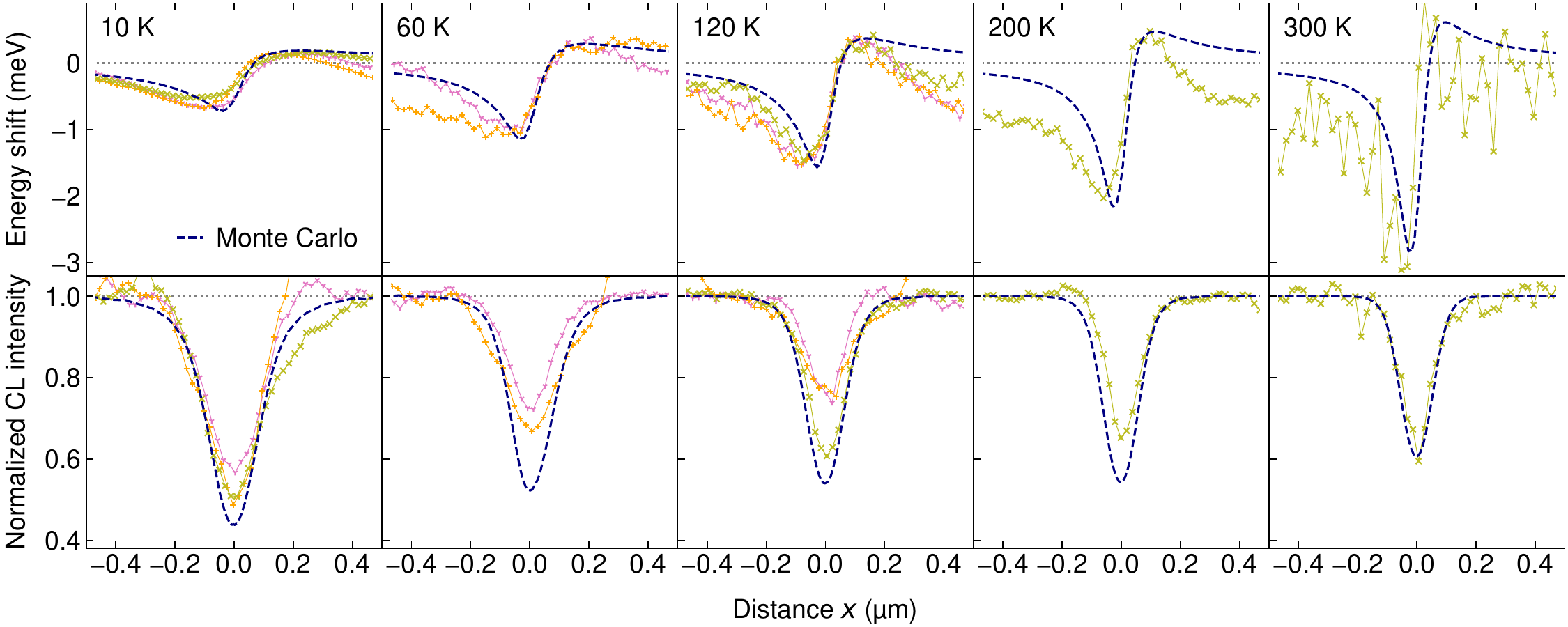}} 
\caption{Profiles of the relative energy shift (top) and normalized CL intensity (bottom) centered on the outcrops of dislocations at the surface recorded at different temperatures for $V=3$~kV. Data for three dislocations, two of which are contained in Fig.~\ref{figMap}, are shown in different colors. The dashed lines show corresponding profiles obtained from Monte Carlo simulations. For the experimental curves, the reference level of the energy shift was chosen to get an optimum match with the maxima and minima of the calculated curves.}
\label{figE}
\end{figure*}

In CD2 \cite{Brandt_2020}, we have shown that the carrier diffusion length in the investigated sample exhibits a clear temperature dependence. Here, we independently examine the relation between carrier diffusion and the CL signal at threading dislocations for the same sample. To this end, we follow the methodology introduced in Ref.~\citenum{Kaganer_2019} for a freestanding GaN layer grown by hydride vapor phase epitaxy with a dislocation density as low as 6$\times 10^5$~cm$^{-2}$. Because the dislocation density of the present sample is three orders of magnitude higher, the strain fields of individual dislocations overlap, creating a rather complex potential landscape. Nevertheless, isolated threading dislocations with sufficiently large defect free surroundings are available to measure both the CL intensity and energy around the outcrop of such single dislocations.

Figures~\ref{figMap}(a) and \ref{figMap}(b) show the maps of the relative energy shift and the intensity extracted from a hyperspectral CL map of the near-band-edge emission of GaN recorded in the area marked in Fig.~\ref{figExp}(b). Automated Lorentzian line-shape fits of the dominant donor-bound exciton line yield the emission energy---shown as relative shift in Fig.~\ref{figMap}(a)---as well as the emission intensity presented in Fig~\ref{figMap}(b) for each pixel of the map. About two dozen dislocations are visible within the latter map, most of which are clustered close to each other. This proximity leads to a partial overlap of the zones of reduced intensity, but more importantly to a quite complex landscape of the emission energy due to the superposition of the strain fields of the individual dislocations \cite{Gmeinwieser_2007}. However, two isolated dislocations, separated from their neighbours by at least 800~nm, are present in the central part of the map. To highlight the dipole-like shift in emission energy at these dislocations, we show excerpts of the relative emission energy map around the two dislocations in Figs.~\ref{figMap}(c) and \ref{figMap}(d). We use this CL map to extract profiles of both the emission energy and intensity aligned along the dipole-like structure of the energy contrast for the two isolated dislocations, as marked by the lines in Figs.~\ref{figMap}(a) and \ref{figMap}(b). The direction of the profiles is chosen to follow the maximum gradient in the energy map, while crossing the center of the intensity dip. The angle between the two dislocations is approximately 120$^\circ$, in accordance with the crystal symmetry. Note that a misalignment of the profile direction by 5$^\circ$ results in an error for the energy shift of about 7\%, which is included in the error estimate. To improve the signal-to-noise ratio, the signal is integrated over the width of the stripe indicated by dashed lines.

For the same two dislocations, as well as for a third one not contained in the map in Fig.~\ref{figMap}, we recorded hyperspectral maps for temperatures up to 300~K. The resulting profiles of emission energy and intensity are presented in Fig.~\ref{figE} and compared with Monte Carlo simulations based on the carrier drift and diffusion around an edge threading dislocation (dashed lines). Full two-dimensional maps of the simulated CL emission energy and intensity for an isolated dislocation can be found in Ref.~\cite{Kaganer_2019} for comparison.

The energy profiles exhibit a dipole-like shift around the dislocation, clearly indicating that we are dealing with a strain distribution induced by either $a$-type (edge) or $a+c$-type (mixed) threading dislocations (the energy shift resulting from the shear strain induced by the screw component is too small to distinguish between the two types \cite{Gmeinwieser_2005}). 
The observed line shift $\Delta E$ between the maxima and minima in Fig.~\ref{figE} amounts to 0.7~meV at 10~K and increases with temperature up to 3.6~meV at 300~K. This shift is remarkably similar among the three investigated dislocations. The increased noise for higher temperatures is due to the thermal broadening of the emission lines from 3~meV at 10~K to 60~meV at 300~K as well as a decreasing intensity, both contributing to a reduced accuracy of the spectral position obtained from the line-shape fit.

Concerning the intensity profiles, their width decreases by about 40\% from 10 to 300~K. At the same time, the depth of the intensity dip centered on the dislocation (hereafter referred to as intensity contrast) decreases, but contrary to the energy variation, the intensity contrast varies significantly between the individual dislocations. The moderate decrease in width is fully consistent with the temperature-dependence of the spatial extent of the CL generation volume as determined in CD1 \cite{Jahn_2020}. Note that the diffusion length for this sample has been found in CD2 \cite{Brandt_2020} to decrease by a factor of 5 between 10 and 300~K. This change in diffusion length is obviously not reflected in the width of the CL intensity profile.

\subsection{Diffusion length from CL maps}
\label{sec:ldiff}

Based on the interpretation of dislocations as nonradiative carrier sinks, various studies have aimed at deducing the diffusion length $L$ in GaN from the width of the CL intensity profile around threading dislocations \cite{Rosner_1997, Sugahara_1998, Cherns_2001, Nakaji_2005, pauc_2006a, Ino_2008}. The resulting values cluster at around 100\,--\,200~nm independent of the sample quality or the experimental conditions, such as the temperature. This fact casts serious doubts on these results, since we would expect $L$ to depend sensitively on crystal quality, temperature, doping, etc. In fact, we can identify two conceptual deficiencies in the approach used in these previous studies. First, in the majority of these studies, the diffusion length has been extracted using an inappropriate approximation of the actual solution of the corresponding diffusion problem \cite{Yakimov_2010, Yakimov_2015, Sabelfeld_2017}. Second, and even more important, physical effects such as exciton drift in the strain field and exciton dissociation in the piezoelectric field of the dislocation have been neglected altogether \cite{Kaganer_2018}. When these effects are properly taken into account, it becomes clear that the diffusion length cannot be extracted from the CL intensity contrast around dislocations, despite the intuitive appeal of this method \cite{Kaganer_2019}. The results shown in the bottom row of Fig.~\ref{figE} confirm this theoretical prediction, since the width of the profile changes only moderately with temperature, while $L$ varies by a factor of 5. It is thus no surprise that studies based on an analysis of the CL intensity contrast have not yielded a systematic temperature dependence for $L$ \cite{Rosner_1997, Sugahara_1998, Cherns_2001, Nakaji_2005, pauc_2006a, Ino_2008}.

In contrast, the top row of Fig.~\ref{figE} shows that the energy shift around the dislocation exhibits a strong dependence on temperature. This effect is easy to understand: the larger the diffusion length, the higher is the fraction of excitons recombining far from the dislocation with an energy close to that of the unstrained bulk and contributing to the detected emission spectrum for a specific excitation position. Effectively, diffusion thus smooths out the strain dipole around the edge threading dislocation  \cite{Kaganer_2019}. Hence, with appropriate modeling, the energy shift facilitates a determination of the diffusion length.

To do so, the CL energy and intensity profiles are simulated using a comprehensive, three-dimensional Monte Carlo approach based on a classical drift-diffusion model \cite{Sabelfeld_1991, Sabelfeld_2016, Kaganer_2019} taking into account the temperature-dependent lateral distribution of the generated carriers as determined in CD1 \cite{Jahn_2020}, the carrier lifetimes measured in CD2 \cite{Brandt_2020}, and the piezoelectric field around the dislocation as derived in Ref.~\onlinecite{Kaganer_2018}.

The strain around the dislocation is modeled in the framework of the elastic continuum approximation, which predicts a diverging strain and thus band gap energy close to the dislocation core. An atomistic treatment of the dislocation shows that the strain and thus band gap variation remain of course finite \cite{Boleininger_2018}. In the present context, the deviation of the continuum approximation close to the dislocation core is of no importance, since the energy variation is smoothed out by the finite generation volume \cite{Gmeinwieser_2007}. Even for the lowest beam energy used here, the resulting energy profile is flatter than the one predicted by atomistic approaches. Note that the pronounced asymmetry of the energy profiles originates from the shear components of the strain tensor \cite{Kaganer_2019}.

At the same time, the theoretical energy profile remains significantly steeper than the profiles observed experimentally (see Fig.~\ref{figE}), even at 300~K, and the (weak) temperature-dependence of the generation volume cannot account for the strong variation of the energy shift in the experiment. In fact, the CL generation volume for $V=3$~kV is very much smaller than in the PL measurements of Ref.~\citenum{Gmeinwieser_2007}, and diffusion thus becomes the dominant process smoothing this profile. In consequence, we can use the mentioned dependence of the energy shift on the diffusion length to determine the diffusion length.

The simulations with $L$ as the only free parameter varied in order to obtain the best possible agreement with the experiments are displayed in Fig.~\ref{figE} \footnote{Note that for a nominal temperature of 10~K, the high-energy slope of the CL spectra indicates a carrier temperature of 20--30~K. Thus, as in Ref.~\citenum{Kaganer_2019}, we have chosen $T=20$~K for the calculated profiles.}. Considering the complexity of the problem, and the fact that we use only $L$ as adjustable parameter, the agreement for the energy shift is very good and still satisfactory for the intensity contrast. While the energy shift is quite robust against any other parameter variations apart from the diffusion length, the intensity contrast strongly depends on several parameters that are not well known. These parameters include the shear component $e_{15}$ of the piezotensor and the Debye screening length associated to the background doping in the sample \cite{Kaganer_2018}. Note that, for the moment, we do not take into account a possible nonradiative recombination at the dislocation line as it has only a negligible influence for low values of $V$.

\begin{figure} 
\centering{\includegraphics[width=1\columnwidth]{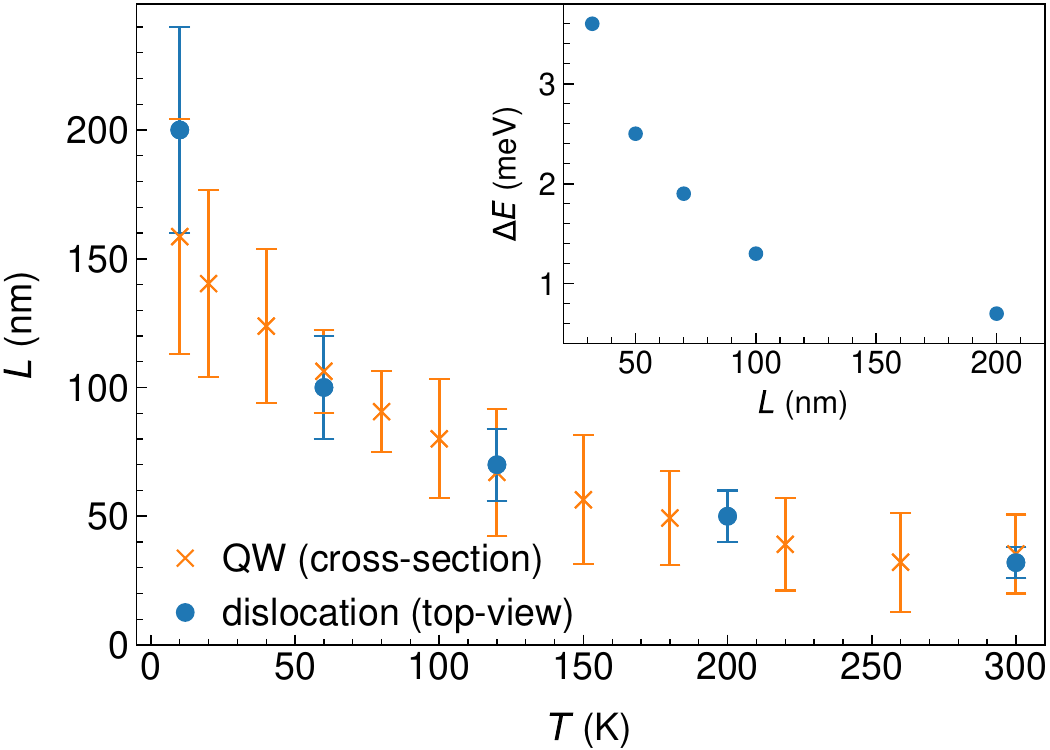}} 
\caption{Diffusion length $L$ as a function of temperature $T$. The circles show the values of $L$ used to simulate the energy profiles in Fig.~\ref{figE}; error bars are $\pm20$\%. The crosses represent the values of $L$ determined in CD2 from profiles of the QW emission excited on the cross-section of the same sample. Here, the error bars correspond to three times the standard deviation of the mean of several measurements. The inset illustrates the relation of the shift in emission energy ($\Delta E$) at the dislocation and the diffusion length.}
\label{figL}
\end{figure}

The resulting values for $L$ are summarized in Fig.~\ref{figL} and compared to those determined in CD2 \cite{Brandt_2020} using the same sample in cross-section employing the (In,Ga)N QW as radiative carrier sink. The values agree perfectly down to $T=60$~K. Only at 10~K, a larger diffusion length is obtained from the energy profiles at dislocations, but the  difference is still within the error bars of the two measurements. For the present measurements, the error bar actually increases with $L$, since the influence of $L$ on the energy shift becomes weaker with increasing $L$. For example, an increase of $L$ from 150 to 200~nm results in a decrease of $\Delta E$ by only 0.1~meV (see also inset to Fig.~\ref{figL}). Therefore, diffusion lengths significantly larger than 200~nm cannot be determined reliably with this method. In contrast, the approach is very well suited for the detection of diffusion lengths $\sigma/2 < L < 200$~nm, with the lower limit given by the spatial extent $\sigma$ of the generation volume determined in CD1 \cite{Jahn_2020}. For room temperature, we will thus not be able to detect a diffusion length less than 10~nm for an acceleration voltage of 3~kV.

\subsection{Anisotropy of the CL image at room temperature}\label{appendix:anisotropy}

\begin{figure}
\includegraphics[width=0.9\columnwidth]{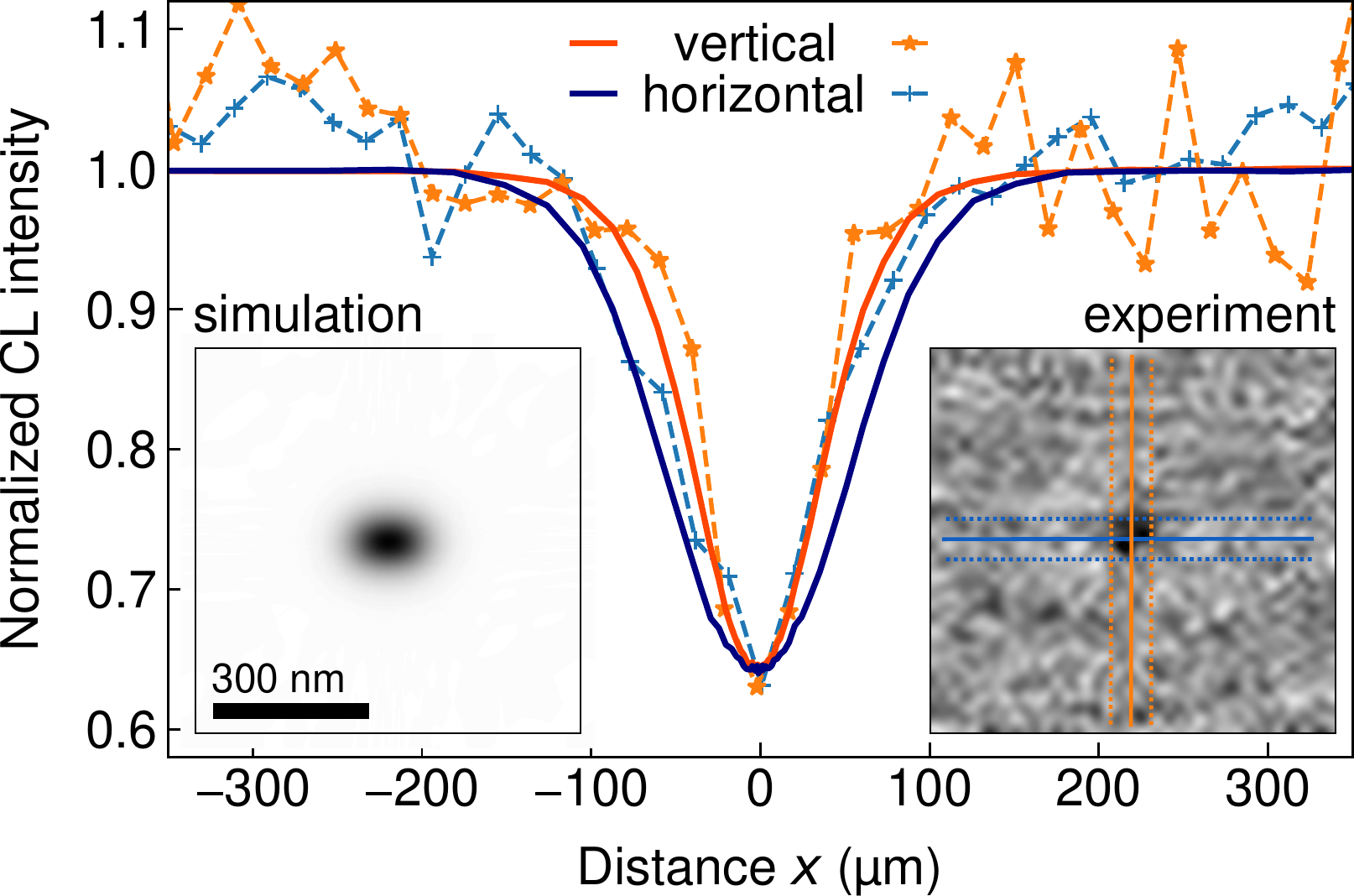}
	\caption{Theoretical (solid line) and experimental (symbols and dashed line) intensity profiles of the GaN CL across a dislocation at room temperature. The energy dipole is oriented along the horizontal direction, for which the intensity profiles are slightly wider. The insets show the simulated and experimental 2D maps of the emission intensity from which the profiles have been taken. For the latter, the profiles were obtained by integrating over the respective intervals indicated by the lines.}
\label{figaniso} 
\end{figure}

A consequence of the short diffusion length at room temperature is that in CL intensity maps the dark spot around a dislocation exhibiting a piezoelectric field should appear ellipsoidal instead of perfectly circular. As shown in Ref.~\cite{Kaganer_2019}, the anisotropy of the emission intensity expected around threading dislocations with an edge component \cite{Kaganer_2018}---and thus with a piezofield close to the surface---is smoothed out by carrier diffusion. Therefore, the dark spot in the CL maps appears to be perfectly round for a sufficiently large diffusion length. At room temperature, at which the diffusion length amounts to only 30~nm, a slight anisotropy of the dark spot remains in a simulated CL intensity map as shown in the left inset of Fig.~\ref{figaniso}. Profiles of the GaN emission intensity across the dislocation show a corresponding difference in width along (horizontal) and perpendicular to (vertical) the orientation of the energy dipole (Fig.~\ref{figaniso}). Experimental maps recorded at room temperature  (right inset of Fig.~\ref{figaniso}) suffer from a limited signal-to-noise ratio, but profiles extracted from such maps exhibit an anisotropy consistent with the theoretical prediction (Fig.~\ref{figaniso}). This result constitutes an additional demonstration of the impact of the piezoelectric field for CL maps of GaN(0001) and of the capabilities of our theoretical model.

\subsection{Depth dependence of the CL intensity contrast} 
\label{sec:depth} 

\begin{figure*} 
\includegraphics[width=1\textwidth]{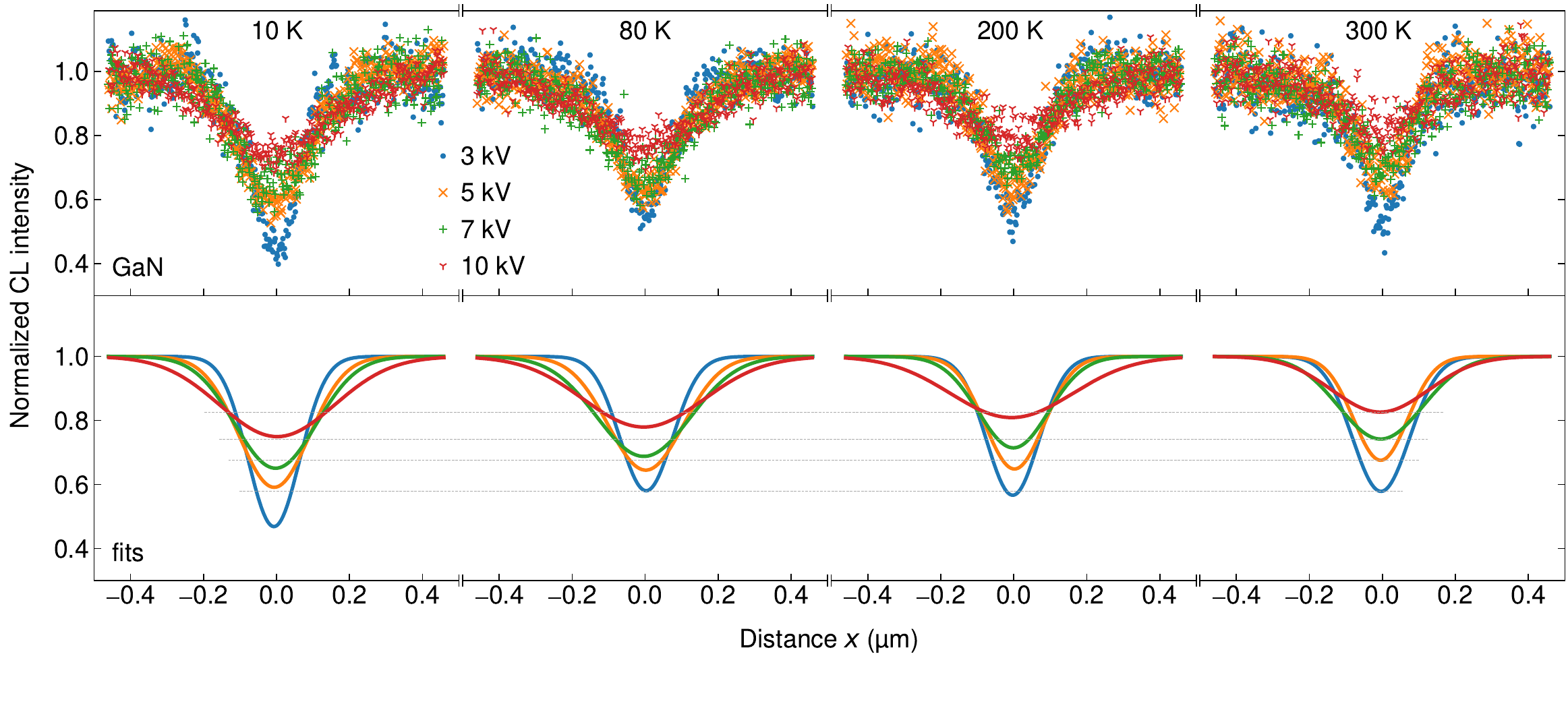} 
	\caption{Profiles of the normalized CL intensity across a dislocation extracted from monochromatic CL maps for acceleration voltages between 3 and 10~kV and temperatures from 10 to 300~K as indicated in the figure. The upper panel presents the experimental data, while the lower panel shows corresponding Gaussian fits and some guides to the eye (dashed horizontal lines) to clearly visualize the trend of the intensity contrast with temperature for a given voltage.}
	\label{figVacc} 
\end{figure*}

For the preceding study of the diffusion length, we have worked with low acceleration voltages, where the carriers are generated within the reach of the piezoelectric field surrounding the outcrop of the dislocation at the surface. Acceleration voltages of $V \leq 5$~kV are commonly employed in scanning electron microscopy-based CL measurements as they provide the highest spatial resolution by minimizing the generation volume. However, under these conditions, the dark contrast at the dislocations can be explained even without invoking nonradiative recombination at the dislocation line. Hence, the common understanding concerning the nonradiative nature of dislocations with an edge component in GaN \cite{Sugahara_1998, Cherns_2001, Miyajima_2001, Yamamoto_2003, Gmeinwieser_2005, Liu_2016} has to be reconsidered. In the following, we will reexamine the effect of the dislocation line on carrier recombination. To this end, we will employ measurements as a function of acceleration voltage and thus excitation depth to probe the sample beyond the region affected by the piezoelectric field. Instead of recording full hyperspectral maps, we now rely on CL monochromatic photon counting maps that can be obtained much more rapidly in analogy to the profiles on the cross-section of a QW presented in CD1 \cite{Jahn_2020} and CD2 \cite{Brandt_2020}. 

Figure~\ref{figVacc} displays such profiles of one dislocation for different acceleration voltages and temperatures. The detection window encompasses the energy range of the near-band-edge emission of GaN, and the profiles are always extracted along the same direction. At all temperatures, the CL intensity contrast decreases with increasing acceleration voltage. This effect is most pronounced at 10~K, where we also observe the strongest absolute intensity contrast. The reduction of this contrast with increasing temperature is monotonic at a fixed acceleration voltage.

The increasing contrast with decreasing temperature would be difficult to understand without the simulations shown in Fig.~\ref{figE}. In fact, both the diffusion length and the generation volume become larger when the temperature is reduced, which should lead to a \emph{decreasing} intensity contrast as the contribution of carriers recombining away from the dislocation is enhanced. However, this effect is overcompensated by the increasing influence of exciton drift in the strain field at the dislocation outcrop with decreasing temperature \cite{Kaganer_2019}. The impact of drift is strongest at 3~kV, where the generation volume does not extend beyond the reach of this strain field, which explains why the temperature dependence is strongest at this acceleration voltage.

The reduced contrast for higher acceleration voltages is a direct consequence of the larger probe volume and the fact that an increasing portion of carriers recombines far from the dislocation. However, this argument applies to both the piezoelectric field close to the surface as well as to the dislocation line itself. As we increase the acceleration voltage, the electrons are scattered to a greater depth beyond the reach of the piezoelectric field, but their lateral spread also increases as quantified in CD1 \cite{Jahn_2020}. From the experiments presented so far, we cannot distinguish these two effects. Therefore, in the following, we go one step further and examine the dislocation-induced contrast in the emission of the buried QW in order to resolve the question whether the dislocation line is nonradiatively active or not.

\subsection{Nonradiative activity of dislocations} 
\label{sec:qw} 

\begin{figure*}
\includegraphics[width=1\textwidth]{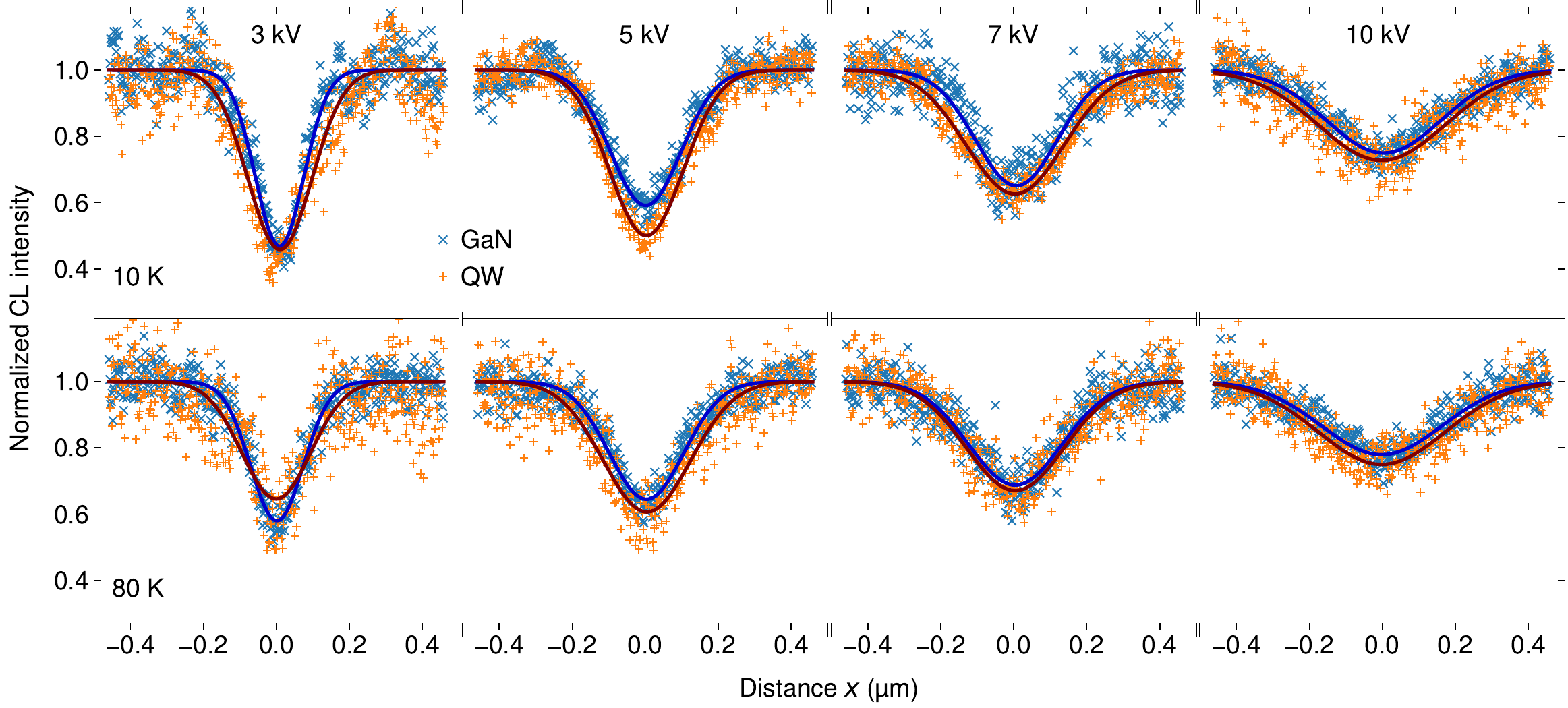} 
	\caption{Comparison of CL intensity profiles across a dislocation obtained from monochromatic images of both the GaN emission at 3.48 ($\times$) and the (In,Ga)N QW emission centered at 2.7~eV ($+$) for different temperatures and acceleration voltages as indicated in the figure. Gaussian fits to the data (solid lines) help to visualize the small, but systematic differences in intensity contrast between the GaN and the QW emission.} 
\label{figQW} 
\end{figure*}

Figure~\ref{figQW} compares the CL intensity profiles obtained across the same dislocation for the emission of the GaN matrix and of the buried (In,Ga)N QW, respectively, at $T=10$ and 80~K and for $V$ ranging from 3 to 10~kV. At the first glance, the experimental intensity profiles for the GaN and the QW emission seem to coincide for all measurement conditions. However, Gaussian fits to the data indicate that the CL intensity contrast is systematically higher for the QW emission except for the lowest acceleration voltage (3~kV). As a first step toward an understanding of these profiles, it is important to recall that in CL spectroscopy excitation is local, but detection is integral. The nonradiative recombination induced by the piezoelectric field at the surface or by the dislocation line in the GaN matrix is thus also reflected in the QW signal. In contrast, the dislocation line in the QW itself has no directly detectable influence on the QW signal, since only an insignificant fraction of carriers will actually reach the QW at the position of the dislocation after traversing a distance of up to 650~nm by diffusion. On first sight, it thus might seem plausible that the intensity profiles for GaN and the QW are identical. However, on closer examination, the mechanisms contributing to the GaN and QW CL contrasts are actually different. 

To understand the profiles quantitatively, the buried QW is integrated into our Monte Carlo drift-diffusion model as perfect (radiative) sink for carriers, i.\,e., we assume that any trajectory reaching the QW plane counts toward the QW emission. Thus, we can concurrently simulate the CL intensity profiles for emission from the top GaN layer and the QW. Again, the experimentally determined generation volume from CD1 \cite{Jahn_2020}, the diffusion lengths obtained above, and the lifetimes from CD2 \cite{Brandt_2020} are used in the simulations. For the GaN signal, the depth-dependent reabsorption of the emission in the layer is additionally taken into account.

\begin{figure*}
\includegraphics[width=0.8\textwidth]{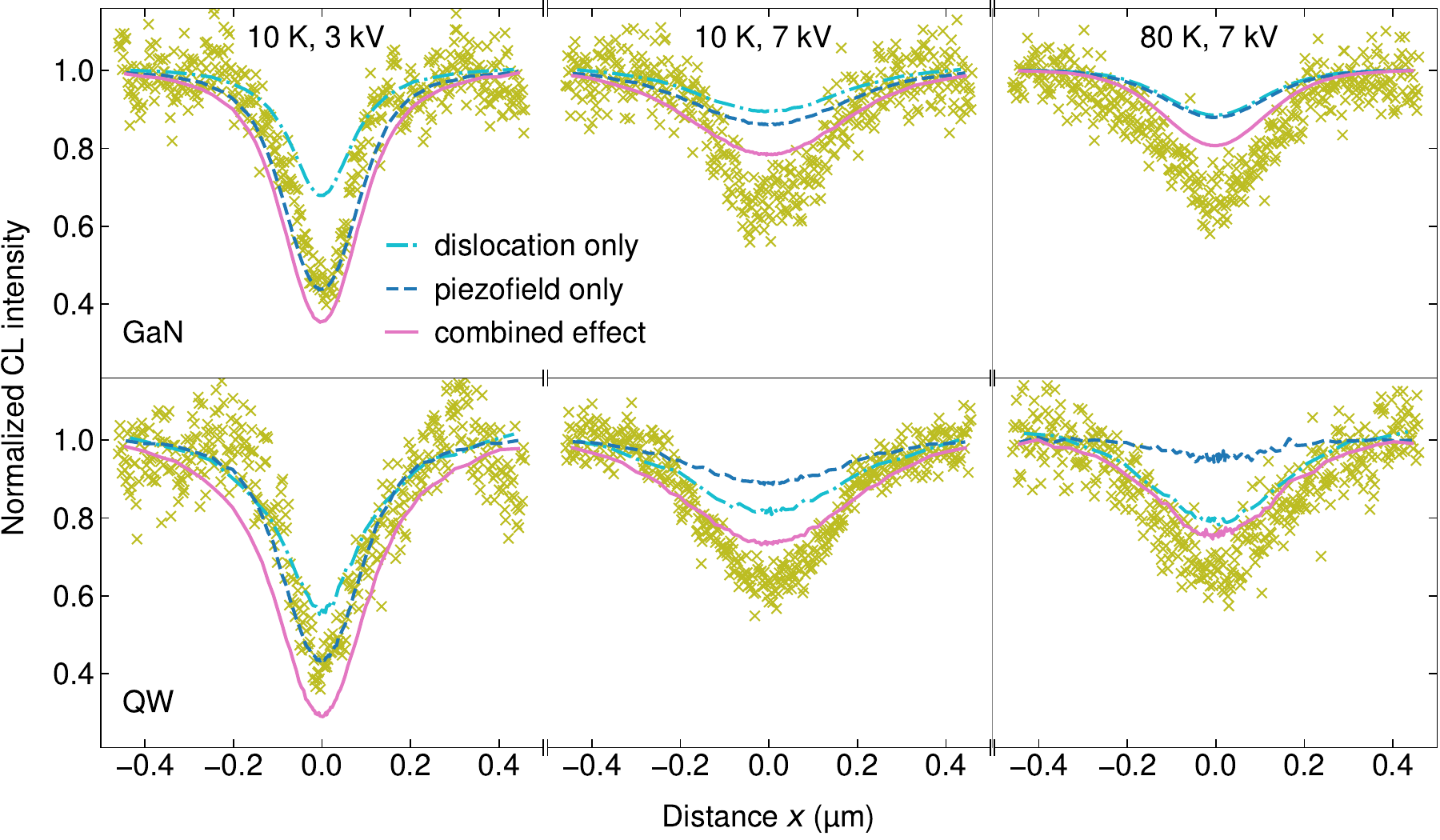} 
	\caption{Experimental (symbols) and simulated (lines) profiles of the CL intensity across a dislocation at different temperatures and acceleration voltages for both the GaN emission (top) and the QW emission (bottom). The dash-dotted lines show results of Monte Carlo simulations when ignoring the effect of the piezoelectic field by treating the dislocation as a 3-nm-thick cylinder with infinite recombination velocity. The dashed lines take into account drift, diffusion and the exciton dissociation in the piezoelectric field of the dislocation, while nonradiative recombination at the dislocation line is ignored. For the solid lines, both the piezoelectric field and nonradiative recombination at the dislocation line are included in the simulations.}
\label{figMC} 
\end{figure*}

In Figure~\ref{figMC}, simulated profiles of the CL intensity, both for the GaN and the QW emission, are compared to a selection of experimental profiles measured across the dislocation. Three different situations are simulated. First, as above, only the nonradiative recombination induced by the piezoelectric field near the surface is considered, and the recombination velocity at the dislocation line is set to zero. Second, only the nonradiative recombination at the dislocation line is taken into account, with the dislocation line being represented by a thin cylinder with infinite recombination velocity. Third, both effects are included in the simulation. Note that both, the experimental and simulated profiles, are always normalized to the intensity far from the dislocation to highlight the intensity contrast. A lower contrast should not be confused with a higher absolute emission intensity.

As expected, the simulated curves at $V=3$~kV, where the generation volume does not extend beyond the reach of the piezoelectric field, are only weakly affected by nonradiative recombination at the dislocation line. However, for an acceleration voltage of 7~kV and especially at $T=80$~K, the difference between the simulated curves with active or inactive dislocation line becomes more pronounced, particularly for the QW emission. Here, it is of importance that the QW is not directly excited by the scattered electrons, but only by carriers diffusing from the generation volume to the QW. Thus, carriers have to traverse up to 650~nm to contribute to the QW emission, while the GaN emission originates from shallower depths.

Simulations taking into account only the active dislocation line, but ignoring the piezoelectric field, exhibit an intensity contrast that is invariably lower than the actual one. This effect is particularly clear for the GaN emission. Note that the dislocation has been assumed to exhibit infinite recombination velocity, and thus has the maximum possible effect on the intensity contrast.

The simulations taking into account only the piezoelectric field, but ignoring the active dislocation line, show a strong intensity contrast close to the surface (for an acceleration voltage of 3 kV), but a very weak one far from it (at 7 kV). In fact, at 7 kV, a large portion of carriers is generated beyond the reach of this field. Due to the shorter diffusion path (average trajectory length) compared with carriers generated closer to the surface, these carriers contribute more significantly to the QW emission. As a result, the QW emission exhibits a lower intensity contrast compared with the GaN emission. This difference is even more pronounced for a smaller diffusion length. Thus, at $T=80$~K, the simulated profile for the QW is almost flat, which implies that even far from the piezoelectric field only few carriers reach the QW. The small diffusion length at 80~K thus leads to a weak QW signal, and we had to increase the number of trajectories by a factor of 10 compared to the corresponding simulations at 10~K to obtain an adequate statistics, while still having a higher noise level in the simulated profiles. In fact, for $T > 80$~K, the diffusion length becomes too small to lead to a detectable CL signal of the buried QW at the investigated acceleration voltages, both theoretically and experimentally.

The simulations with both piezoelectric field and active dislocation line (solid lines in Fig.~\ref{figMC}) show the opposite behavior. The intensity contrast is higher for the QW compared with the GaN emission even though the contribution from the piezoelectric field is reduced. This higher sensitivity of the QW emission for the impact of the dislocation line is again due to the long diffusion path: carriers reaching the QW plane have a higher chance to interact with the dislocation line. This path becomes even longer when the diffusion length is reduced at $T=80$~K.

Overall, the experimental intensity contrast can only be explained by taking into account the piezoelectric field close to the surface and at the same time assuming that the dislocation line acts as efficient nonradiative recombination channel. For the deeply buried QW, the latter is the major contribution to the strong intensity contrast. This result demonstrates that the emission from the underlying QW is an excellent probe for the nonradiative activity of the dislocation in the bulk of the GaN layer up to a well-defined depth of 650~nm. Further implications for buried heterostructures, such as used in light emitting devices, are (i) that the dislocation should still act as strong nonradiative center in the quantum wells and (ii) that the impact of the dislocation line on the luminous efficiency is then controlled by the diffusion length in the active region.

\section{Summary and conclusion}
\label{sec:summary}

In this work, we have presented a comprehensive investigation and critical assessment of the CL signal around threading dislocations in a custom-designed heterostructure consisting of a thick GaN(0001) layer with a deeply buried (In,Ga)N quantum well. The motivation for our scrutiny was the recognition that the understanding of the impact of threading dislocations in GaN on carrier recombination is incomplete at best, and possibly even grossly mistaken. Our study demonstrates that the nonradiative nature of edge or mixed type dislocations in GaN has a twofold origin depending on the distance from the surface. Close to the surface, it is dominated by the piezoelectric field surrounding the dislocation outcrop, which is the cause of the dark spots observed in maps of the CL intensity. Far from the surface, the nonradiative activity of the dislocation line itself dominates. In heterostructures with an active region buried at a depth significantly larger than $140$~nm, such as used in light emitting devices, the impact of the dislocation line on the luminous efficiency is then controlled by the diffusion length in the active region.

In previous studies, the spatial extent of the dark spots in CL maps was believed to be determined by carrier diffusion. In fact, most of the values available for the carrier diffusion length in GaN were obtained in this intuitively appealing and experimentally straightforward way. However, our investigation shows that the spot size is hardly influenced by temperature, and is in fact rather controlled by the spatial extent of the piezoelectric field, and to a lesser degree by the lateral width of the CL generation volume. We have determined the actual diffusion length from the dipole-like energy shift of the emission at the dislocation outcrop, and found that it varies by a factor of 5 between temperatures of 10 and 300~K, in excellent agreement with the values derived independently in CD2 from CL intensity profiles across the buried quantum well. This agreement supports the use of the energy shift across a dislocation for determining the diffusion length by cathodoluminescence measurements. Our results underline the need for new studies of the diffusion length in GaN. Particularly valuable would be a systematic variation of temperature at different doping densities, as well as a comparison of samples obtained by different growth methods to understand the impact of point defects on the carrier lifetime and thus the diffusion length.

\begin{acknowledgments}
The authors are indebted to Manfred Ramsteiner for a critical reading of the manuscript. Special thanks are due to Holger Grahn, Lutz Geelhaar, Achim Trampert and Henning Riechert for their continuous encouragement and support. K.~K.~S.\ and A.~E.~K.\ acknowledge funding from the Russian Science Foundation under grant N 19-11-00019.
\end{acknowledgments}



\bibliography{references} 

\begin{thebibliography}{67}%
\makeatletter
\providecommand \@ifxundefined [1]{%
 \@ifx{#1\undefined}
}%
\providecommand \@ifnum [1]{%
 \ifnum #1\expandafter \@firstoftwo
 \else \expandafter \@secondoftwo
 \fi
}%
\providecommand \@ifx [1]{%
 \ifx #1\expandafter \@firstoftwo
 \else \expandafter \@secondoftwo
 \fi
}%
\providecommand \natexlab [1]{#1}%
\providecommand \enquote  [1]{``#1''}%
\providecommand \bibnamefont  [1]{#1}%
\providecommand \bibfnamefont [1]{#1}%
\providecommand \citenamefont [1]{#1}%
\providecommand \href@noop [0]{\@secondoftwo}%
\providecommand \href [0]{\begingroup \@sanitize@url \@href}%
\providecommand \@href[1]{\@@startlink{#1}\@@href}%
\providecommand \@@href[1]{\endgroup#1\@@endlink}%
\providecommand \@sanitize@url [0]{\catcode `\\12\catcode `\$12\catcode
  `\&12\catcode `\#12\catcode `\^12\catcode `\_12\catcode `\%12\relax}%
\providecommand \@@startlink[1]{}%
\providecommand \@@endlink[0]{}%
\providecommand \url  [0]{\begingroup\@sanitize@url \@url }%
\providecommand \@url [1]{\endgroup\@href {#1}{\urlprefix }}%
\providecommand \urlprefix  [0]{URL }%
\providecommand \Eprint [0]{\href }%
\providecommand \doibase [0]{https://doi.org/}%
\providecommand \selectlanguage [0]{\@gobble}%
\providecommand \bibinfo  [0]{\@secondoftwo}%
\providecommand \bibfield  [0]{\@secondoftwo}%
\providecommand \translation [1]{[#1]}%
\providecommand \BibitemOpen [0]{}%
\providecommand \bibitemStop [0]{}%
\providecommand \bibitemNoStop [0]{.\EOS\space}%
\providecommand \EOS [0]{\spacefactor3000\relax}%
\providecommand \BibitemShut  [1]{\csname bibitem#1\endcsname}%
\let\auto@bib@innerbib\@empty
\bibitem [{\citenamefont {Rosner}\ \emph {et~al.}(1997)\citenamefont {Rosner},
  \citenamefont {Carr}, \citenamefont {Ludowise}, \citenamefont {Girolami},\
  and\ \citenamefont {Erikson}}]{Rosner_1997}%
  \BibitemOpen
  \bibfield  {author} {\bibinfo {author} {\bibfnamefont {S.~J.}\ \bibnamefont
  {Rosner}}, \bibinfo {author} {\bibfnamefont {E.~C.}\ \bibnamefont {Carr}},
  \bibinfo {author} {\bibfnamefont {M.~J.}\ \bibnamefont {Ludowise}}, \bibinfo
  {author} {\bibfnamefont {G.}~\bibnamefont {Girolami}},\ and\ \bibinfo
  {author} {\bibfnamefont {H.~I.}\ \bibnamefont {Erikson}},\ }\bibfield
  {title} {\bibinfo {title} {Correlation of cathodoluminescence inhomogeneity
  with microstructural defects in epitaxial {{GaN}} grown by metalorganic
  chemical-vapor deposition},\ }\href {https://doi.org/10.1063/1.118322}
  {\bibfield  {journal} {\bibinfo  {journal} {Appl. Phys. Lett.}\ }\textbf
  {\bibinfo {volume} {70}},\ \bibinfo {pages} {420} (\bibinfo {year}
  {1997})}\BibitemShut {NoStop}%
\bibitem [{\citenamefont {Hangleiter}\ \emph {et~al.}(2005)\citenamefont
  {Hangleiter}, \citenamefont {Hitzel}, \citenamefont {Netzel}, \citenamefont
  {Fuhrmann}, \citenamefont {Rossow}, \citenamefont {Ade},\ and\ \citenamefont
  {Hinze}}]{Hangleiter_2005}%
  \BibitemOpen
  \bibfield  {author} {\bibinfo {author} {\bibfnamefont {A.}~\bibnamefont
  {Hangleiter}}, \bibinfo {author} {\bibfnamefont {F.}~\bibnamefont {Hitzel}},
  \bibinfo {author} {\bibfnamefont {C.}~\bibnamefont {Netzel}}, \bibinfo
  {author} {\bibfnamefont {D.}~\bibnamefont {Fuhrmann}}, \bibinfo {author}
  {\bibfnamefont {U.}~\bibnamefont {Rossow}}, \bibinfo {author} {\bibfnamefont
  {G.}~\bibnamefont {Ade}},\ and\ \bibinfo {author} {\bibfnamefont
  {P.}~\bibnamefont {Hinze}},\ }\bibfield  {title} {\bibinfo {title}
  {Suppression of nonradiative recombination by {{V}}-shaped pits in
  {{GaInN}}/{{GaN}} quantum wells produces a large increase in the light
  emission efficiency},\ }\href {https://doi.org/10.1103/PhysRevLett.95.127402}
  {\bibfield  {journal} {\bibinfo  {journal} {Phys. Rev. Lett.}\ }\textbf
  {\bibinfo {volume} {95}},\ \bibinfo {pages} {127402} (\bibinfo {year}
  {2005})}\BibitemShut {NoStop}%
\bibitem [{\citenamefont {Chichibu}\ \emph {et~al.}(2006)\citenamefont
  {Chichibu}, \citenamefont {Uedono}, \citenamefont {Onuma}, \citenamefont
  {a~Haskell}, \citenamefont {Chakraborty}, \citenamefont {Koyama},
  \citenamefont {Fini}, \citenamefont {Keller}, \citenamefont {DenBaars},
  \citenamefont {Speck}, \citenamefont {Mishra}, \citenamefont {Nakamura},
  \citenamefont {Yamaguchi}, \citenamefont {Kamiyama}, \citenamefont {Amano},
  \citenamefont {Akasaki}, \citenamefont {Han},\ and\ \citenamefont
  {Sota}}]{Chichibu_2006}%
  \BibitemOpen
  \bibfield  {author} {\bibinfo {author} {\bibfnamefont {S.~F.}\ \bibnamefont
  {Chichibu}}, \bibinfo {author} {\bibfnamefont {A.}~\bibnamefont {Uedono}},
  \bibinfo {author} {\bibfnamefont {T.}~\bibnamefont {Onuma}}, \bibinfo
  {author} {\bibfnamefont {B.}~\bibnamefont {a~Haskell}}, \bibinfo {author}
  {\bibfnamefont {A.}~\bibnamefont {Chakraborty}}, \bibinfo {author}
  {\bibfnamefont {T.}~\bibnamefont {Koyama}}, \bibinfo {author} {\bibfnamefont
  {P.~T.}\ \bibnamefont {Fini}}, \bibinfo {author} {\bibfnamefont
  {S.}~\bibnamefont {Keller}}, \bibinfo {author} {\bibfnamefont {S.~P.}\
  \bibnamefont {DenBaars}}, \bibinfo {author} {\bibfnamefont {J.~S.}\
  \bibnamefont {Speck}}, \bibinfo {author} {\bibfnamefont {U.~K.}\ \bibnamefont
  {Mishra}}, \bibinfo {author} {\bibfnamefont {S.}~\bibnamefont {Nakamura}},
  \bibinfo {author} {\bibfnamefont {S.}~\bibnamefont {Yamaguchi}}, \bibinfo
  {author} {\bibfnamefont {S.}~\bibnamefont {Kamiyama}}, \bibinfo {author}
  {\bibfnamefont {H.}~\bibnamefont {Amano}}, \bibinfo {author} {\bibfnamefont
  {I.}~\bibnamefont {Akasaki}}, \bibinfo {author} {\bibfnamefont
  {J.}~\bibnamefont {Han}},\ and\ \bibinfo {author} {\bibfnamefont
  {T.}~\bibnamefont {Sota}},\ }\bibfield  {title} {\bibinfo {title} {Origin of
  defect-insensitive emission probability in {{In}}-containing
  ({{Al}},{{In}},{{Ga}}){{N}} alloy semiconductors.},\ }\href
  {https://doi.org/10.1038/nmat1726} {\bibfield  {journal} {\bibinfo  {journal}
  {Nat. Mater.}\ }\textbf {\bibinfo {volume} {5}},\ \bibinfo {pages} {810}
  (\bibinfo {year} {2006})}\BibitemShut {NoStop}%
\bibitem [{\citenamefont {Oliver}\ \emph {et~al.}(2010)\citenamefont {Oliver},
  \citenamefont {Bennett}, \citenamefont {Zhu}, \citenamefont {Beesley},
  \citenamefont {Kappers}, \citenamefont {Saxena}, \citenamefont {Cerezo},\
  and\ \citenamefont {Humphreys}}]{Oliver_2010}%
  \BibitemOpen
  \bibfield  {author} {\bibinfo {author} {\bibfnamefont {R.~A.}\ \bibnamefont
  {Oliver}}, \bibinfo {author} {\bibfnamefont {S.~E.}\ \bibnamefont {Bennett}},
  \bibinfo {author} {\bibfnamefont {T.}~\bibnamefont {Zhu}}, \bibinfo {author}
  {\bibfnamefont {D.~J.}\ \bibnamefont {Beesley}}, \bibinfo {author}
  {\bibfnamefont {M.~J.}\ \bibnamefont {Kappers}}, \bibinfo {author}
  {\bibfnamefont {D.}~\bibnamefont {Saxena}}, \bibinfo {author} {\bibfnamefont
  {A.}~\bibnamefont {Cerezo}},\ and\ \bibinfo {author} {\bibfnamefont {C.~J.}\
  \bibnamefont {Humphreys}},\ }\bibfield  {title} {\bibinfo {title}
  {Microstructural origins of localization in {{InGaN}} quantum wells},\ }\href
  {https://doi.org/10.1088/0022-3727/43/35/354003} {\bibfield  {journal}
  {\bibinfo  {journal} {J. Phys. D: Appl. Phys.}\ }\textbf {\bibinfo {volume}
  {43}},\ \bibinfo {pages} {354003} (\bibinfo {year} {2010})}\BibitemShut
  {NoStop}%
\bibitem [{\citenamefont {Nakamura}(2015)}]{Nakamura_2015}%
  \BibitemOpen
  \bibfield  {author} {\bibinfo {author} {\bibfnamefont {S.}~\bibnamefont
  {Nakamura}},\ }\bibfield  {title} {\bibinfo {title} {Nobel {{Lecture}}:
  Background story of the invention of efficient blue {{InGaN}} light emitting
  diodes},\ }\href {https://doi.org/10.1103/RevModPhys.87.1139} {\bibfield
  {journal} {\bibinfo  {journal} {Rev. Mod. Phys.}\ }\textbf {\bibinfo {volume}
  {87}},\ \bibinfo {pages} {1139} (\bibinfo {year} {2015})}\BibitemShut
  {NoStop}%
\bibitem [{\citenamefont {Zhu}\ and\ \citenamefont
  {Humphreys}(2016)}]{Zhu_2016}%
  \BibitemOpen
  \bibfield  {author} {\bibinfo {author} {\bibfnamefont {D.}~\bibnamefont
  {Zhu}}\ and\ \bibinfo {author} {\bibfnamefont {C.~J.}\ \bibnamefont
  {Humphreys}},\ }\bibfield  {title} {\bibinfo {title} {Solid-state lighting
  based on light emitting diode technology},\ }in\ \href
  {https://doi.org/10.1007/978-3-319-31903-2_5} {\emph {\bibinfo {booktitle}
  {Optics in {{Our Time}}}}},\ \bibinfo {editor} {edited by\ \bibinfo {editor}
  {\bibfnamefont {M.~D.}\ \bibnamefont {{Al-Amri}}}, \bibinfo {editor}
  {\bibfnamefont {M.}~\bibnamefont {{El-Gomati}}},\ and\ \bibinfo {editor}
  {\bibfnamefont {M.~S.}\ \bibnamefont {Zubairy}}}\ (\bibinfo  {publisher}
  {{Springer International Publishing}},\ \bibinfo {address} {{Cham}},\
  \bibinfo {year} {2016})\ pp.\ \bibinfo {pages} {87--118}\BibitemShut
  {NoStop}%
\bibitem [{\citenamefont {Jones}(2000)}]{Jones_2000}%
  \BibitemOpen
  \bibfield  {author} {\bibinfo {author} {\bibfnamefont {R.}~\bibnamefont
  {Jones}},\ }\bibfield  {title} {\bibinfo {title} {Do we really understand
  dislocations in semiconductors?},\ }\href
  {https://doi.org/10.1016/S0921-5107(99)00344-X} {\bibfield  {journal}
  {\bibinfo  {journal} {Mater. Sci. Eng. B}\ }\textbf {\bibinfo {volume}
  {71}},\ \bibinfo {pages} {24} (\bibinfo {year} {2000})}\BibitemShut {NoStop}%
\bibitem [{\citenamefont {Reshchikov}\ and\ \citenamefont {Morko{\c
  c}}(2005)}]{Reshchikov_jap_2005}%
  \BibitemOpen
  \bibfield  {author} {\bibinfo {author} {\bibfnamefont {M.~A.}\ \bibnamefont
  {Reshchikov}}\ and\ \bibinfo {author} {\bibfnamefont {H.}~\bibnamefont
  {Morko{\c c}}},\ }\bibfield  {title} {\bibinfo {title} {Luminescence
  properties of defects in {{GaN}}},\ }\href
  {https://doi.org/10.1063/1.1868059} {\bibfield  {journal} {\bibinfo
  {journal} {J. Appl. Phys.}\ }\textbf {\bibinfo {volume} {97}},\ \bibinfo
  {pages} {061301} (\bibinfo {year} {2005})}\BibitemShut {NoStop}%
\bibitem [{\citenamefont {You}\ and\ \citenamefont {Johnson}(2009)}]{You_2009}%
  \BibitemOpen
  \bibfield  {author} {\bibinfo {author} {\bibfnamefont {J.}~\bibnamefont
  {You}}\ and\ \bibinfo {author} {\bibfnamefont {H.}~\bibnamefont {Johnson}},\
  }\bibfield  {title} {\bibinfo {title} {Effect of {{Dislocations}} on
  {{Electrical}} and {{Optical Properties}} in {{GaAs}} and {{GaN}}},\ }in\
  \href {https://doi.org/10.1016/S0081-1947(09)00003-4} {\emph {\bibinfo
  {booktitle} {Solid {{State Physics}}}}},\ Vol.~\bibinfo {volume} {61}\
  (\bibinfo  {publisher} {{Elsevier}},\ \bibinfo {year} {2009})\ pp.\ \bibinfo
  {pages} {143--261}\BibitemShut {NoStop}%
\bibitem [{\citenamefont {Lester}\ \emph {et~al.}(1995)\citenamefont {Lester},
  \citenamefont {Ponce}, \citenamefont {Craford},\ and\ \citenamefont
  {Steigerwald}}]{Lester_1995}%
  \BibitemOpen
  \bibfield  {author} {\bibinfo {author} {\bibfnamefont {S.~D.}\ \bibnamefont
  {Lester}}, \bibinfo {author} {\bibfnamefont {F.~A.}\ \bibnamefont {Ponce}},
  \bibinfo {author} {\bibfnamefont {M.~G.}\ \bibnamefont {Craford}},\ and\
  \bibinfo {author} {\bibfnamefont {D.~A.}\ \bibnamefont {Steigerwald}},\
  }\bibfield  {title} {\bibinfo {title} {High dislocation densities in high
  efficiency {{GaN}}-based light-emitting diodes},\ }\href
  {https://doi.org/10.1063/1.113252} {\bibfield  {journal} {\bibinfo  {journal}
  {Appl. Phys. Lett.}\ }\textbf {\bibinfo {volume} {66}},\ \bibinfo {pages}
  {1249} (\bibinfo {year} {1995})}\BibitemShut {NoStop}%
\bibitem [{\citenamefont {Speck}\ and\ \citenamefont
  {Rosner}(1999)}]{Speck_1999}%
  \BibitemOpen
  \bibfield  {author} {\bibinfo {author} {\bibfnamefont {J.}~\bibnamefont
  {Speck}}\ and\ \bibinfo {author} {\bibfnamefont {S.}~\bibnamefont {Rosner}},\
  }\bibfield  {title} {\bibinfo {title} {The role of threading dislocations in
  the physical properties of {{GaN}} and its alloys},\ }\href
  {https://doi.org/10.1016/S0921-4526(99)00399-3} {\bibfield  {journal}
  {\bibinfo  {journal} {Physica B: Condensed Matter}\ }\textbf {\bibinfo
  {volume} {273--274}},\ \bibinfo {pages} {24} (\bibinfo {year}
  {1999})}\BibitemShut {NoStop}%
\bibitem [{\citenamefont {Sugahara}\ \emph {et~al.}(1998)\citenamefont
  {Sugahara}, \citenamefont {Sato}, \citenamefont {Hao}, \citenamefont {Naoi},
  \citenamefont {Kurai}, \citenamefont {Tottori}, \citenamefont {Yamashita},
  \citenamefont {Nishino}, \citenamefont {Romano},\ and\ \citenamefont
  {Sakai}}]{Sugahara_1998}%
  \BibitemOpen
  \bibfield  {author} {\bibinfo {author} {\bibfnamefont {T.}~\bibnamefont
  {Sugahara}}, \bibinfo {author} {\bibfnamefont {H.}~\bibnamefont {Sato}},
  \bibinfo {author} {\bibfnamefont {M.}~\bibnamefont {Hao}}, \bibinfo {author}
  {\bibfnamefont {Y.}~\bibnamefont {Naoi}}, \bibinfo {author} {\bibfnamefont
  {S.}~\bibnamefont {Kurai}}, \bibinfo {author} {\bibfnamefont
  {S.}~\bibnamefont {Tottori}}, \bibinfo {author} {\bibfnamefont
  {K.}~\bibnamefont {Yamashita}}, \bibinfo {author} {\bibfnamefont
  {K.}~\bibnamefont {Nishino}}, \bibinfo {author} {\bibfnamefont {L.~T.}\
  \bibnamefont {Romano}},\ and\ \bibinfo {author} {\bibfnamefont
  {S.}~\bibnamefont {Sakai}},\ }\bibfield  {title} {\bibinfo {title} {Direct
  evidence that dislocations are non-radiative recombination centers in
  {{GaN}}},\ }\href {https://doi.org/10.1143/jjap.37.l398} {\bibfield
  {journal} {\bibinfo  {journal} {Jpn. J. Appl. Phys.}\ }\textbf {\bibinfo
  {volume} {37}},\ \bibinfo {pages} {L398} (\bibinfo {year}
  {1998})}\BibitemShut {NoStop}%
\bibitem [{\citenamefont {Bandi{\'c}}\ \emph {et~al.}(2000)\citenamefont
  {Bandi{\'c}}, \citenamefont {Bridger}, \citenamefont {Piquette},\ and\
  \citenamefont {McGill}}]{Bandic_2000}%
  \BibitemOpen
  \bibfield  {author} {\bibinfo {author} {\bibfnamefont {Z.~Z.}\ \bibnamefont
  {Bandi{\'c}}}, \bibinfo {author} {\bibfnamefont {P.~M.}\ \bibnamefont
  {Bridger}}, \bibinfo {author} {\bibfnamefont {E.~C.}\ \bibnamefont
  {Piquette}},\ and\ \bibinfo {author} {\bibfnamefont {T.~C.}\ \bibnamefont
  {McGill}},\ }\bibfield  {title} {\bibinfo {title} {Values of minority carrier
  diffusion lengths and lifetimes in {{GaN}} and their implications for bipolar
  devices},\ }\href {https://doi.org/10.1016/s0038-1101(99)00227-0} {\bibfield
  {journal} {\bibinfo  {journal} {Solid State Electron.}\ }\textbf {\bibinfo
  {volume} {44}},\ \bibinfo {pages} {221} (\bibinfo {year} {2000})}\BibitemShut
  {NoStop}%
\bibitem [{\citenamefont {Chernyak}\ \emph {et~al.}(2001)\citenamefont
  {Chernyak}, \citenamefont {Osinsky},\ and\ \citenamefont
  {Schulte}}]{Chernyak_2001}%
  \BibitemOpen
  \bibfield  {author} {\bibinfo {author} {\bibfnamefont {L.}~\bibnamefont
  {Chernyak}}, \bibinfo {author} {\bibfnamefont {A.}~\bibnamefont {Osinsky}},\
  and\ \bibinfo {author} {\bibfnamefont {A.}~\bibnamefont {Schulte}},\
  }\bibfield  {title} {\bibinfo {title} {Minority carrier transport in {{GaN}}
  and related materials},\ }\href
  {https://doi.org/10.1016/s0038-1101(01)00161-7} {\bibfield  {journal}
  {\bibinfo  {journal} {Solid State Electron.}\ }\textbf {\bibinfo {volume}
  {45}},\ \bibinfo {pages} {1687} (\bibinfo {year} {2001})}\BibitemShut
  {NoStop}%
\bibitem [{\citenamefont {Karpov}\ and\ \citenamefont
  {Makarov}(2003)}]{Karpov_2003}%
  \BibitemOpen
  \bibfield  {author} {\bibinfo {author} {\bibfnamefont {S.~Y.}\ \bibnamefont
  {Karpov}}\ and\ \bibinfo {author} {\bibfnamefont {Y.~N.}\ \bibnamefont
  {Makarov}},\ }\bibfield  {title} {\bibinfo {title} {Dislocation effect on
  light emission efficiency in gallium nitride},\ }\href
  {https://doi.org/10.1063/1.1527225} {\bibfield  {journal} {\bibinfo
  {journal} {Appl. Phys. Lett.}\ }\textbf {\bibinfo {volume} {81}},\ \bibinfo
  {pages} {4721} (\bibinfo {year} {2003})}\BibitemShut {NoStop}%
\bibitem [{\citenamefont {{\v S}{\v c}ajev}\ \emph {et~al.}(2012)\citenamefont
  {{\v S}{\v c}ajev}, \citenamefont {Jara{\v s}i{\=u}nas}, \citenamefont
  {Okur}, \citenamefont {{\"O}zg{\"u}r},\ and\ \citenamefont {Morko{\c
  c}}}]{Scajev_2012}%
  \BibitemOpen
  \bibfield  {author} {\bibinfo {author} {\bibfnamefont {P.}~\bibnamefont {{\v
  S}{\v c}ajev}}, \bibinfo {author} {\bibfnamefont {K.}~\bibnamefont {Jara{\v
  s}i{\=u}nas}}, \bibinfo {author} {\bibfnamefont {S.}~\bibnamefont {Okur}},
  \bibinfo {author} {\bibfnamefont {{\"U}.}~\bibnamefont {{\"O}zg{\"u}r}},\
  and\ \bibinfo {author} {\bibfnamefont {H.}~\bibnamefont {Morko{\c c}}},\
  }\bibfield  {title} {\bibinfo {title} {Carrier dynamics in bulk {{GaN}}},\
  }\href {https://doi.org/10.1063/1.3673851} {\bibfield  {journal} {\bibinfo
  {journal} {J. Appl. Phys.}\ }\textbf {\bibinfo {volume} {111}},\ \bibinfo
  {pages} {023702} (\bibinfo {year} {2012})}\BibitemShut {NoStop}%
\bibitem [{\citenamefont {Sabelfeld}\ \emph {et~al.}(2017)\citenamefont
  {Sabelfeld}, \citenamefont {Kaganer}, \citenamefont {Pf{\"u}ller},\ and\
  \citenamefont {Brandt}}]{Sabelfeld_2017}%
  \BibitemOpen
  \bibfield  {author} {\bibinfo {author} {\bibfnamefont {K.~K.}\ \bibnamefont
  {Sabelfeld}}, \bibinfo {author} {\bibfnamefont {V.~M.}\ \bibnamefont
  {Kaganer}}, \bibinfo {author} {\bibfnamefont {C.}~\bibnamefont
  {Pf{\"u}ller}},\ and\ \bibinfo {author} {\bibfnamefont {O.}~\bibnamefont
  {Brandt}},\ }\bibfield  {title} {\bibinfo {title} {Dislocation contrast in
  cathodoluminescence and electron-beam induced current maps on
  {{GaN}}(0001)},\ }\href {https://doi.org/10.1088/1361-6463/aa85c8} {\bibfield
   {journal} {\bibinfo  {journal} {J. Phys. D: Appl. Phys.}\ }\textbf {\bibinfo
  {volume} {50}},\ \bibinfo {pages} {405101} (\bibinfo {year}
  {2017})}\BibitemShut {NoStop}%
\bibitem [{\citenamefont {Hino}\ \emph {et~al.}(2000)\citenamefont {Hino},
  \citenamefont {Tomiya}, \citenamefont {Miyajima}, \citenamefont {Yanashima},
  \citenamefont {Hashimoto},\ and\ \citenamefont {Ikeda}}]{Hino_2000}%
  \BibitemOpen
  \bibfield  {author} {\bibinfo {author} {\bibfnamefont {T.}~\bibnamefont
  {Hino}}, \bibinfo {author} {\bibfnamefont {S.}~\bibnamefont {Tomiya}},
  \bibinfo {author} {\bibfnamefont {T.}~\bibnamefont {Miyajima}}, \bibinfo
  {author} {\bibfnamefont {K.}~\bibnamefont {Yanashima}}, \bibinfo {author}
  {\bibfnamefont {S.}~\bibnamefont {Hashimoto}},\ and\ \bibinfo {author}
  {\bibfnamefont {M.}~\bibnamefont {Ikeda}},\ }\bibfield  {title} {\bibinfo
  {title} {Characterization of threading dislocations in {{GaN}} epitaxial
  layers},\ }\href {https://doi.org/10.1063/1.126666} {\bibfield  {journal}
  {\bibinfo  {journal} {Appl. Phys. Lett.}\ }\textbf {\bibinfo {volume} {76}},\
  \bibinfo {pages} {3421} (\bibinfo {year} {2000})}\BibitemShut {NoStop}%
\bibitem [{\citenamefont {Albrecht}\ \emph {et~al.}(2008)\citenamefont
  {Albrecht}, \citenamefont {Weyher}, \citenamefont {Lucznik}, \citenamefont
  {Grzegory},\ and\ \citenamefont {Porowski}}]{Albrecht_2008}%
  \BibitemOpen
  \bibfield  {author} {\bibinfo {author} {\bibfnamefont {M.}~\bibnamefont
  {Albrecht}}, \bibinfo {author} {\bibfnamefont {J.~L.}\ \bibnamefont
  {Weyher}}, \bibinfo {author} {\bibfnamefont {B.}~\bibnamefont {Lucznik}},
  \bibinfo {author} {\bibfnamefont {I.}~\bibnamefont {Grzegory}},\ and\
  \bibinfo {author} {\bibfnamefont {S.}~\bibnamefont {Porowski}},\ }\bibfield
  {title} {\bibinfo {title} {Nonradiative recombination at threading
  dislocations in n-type {{GaN}}: Studied by cathodoluminescence and defect
  selective etching},\ }\href {https://doi.org/10.1063/1.2928226} {\bibfield
  {journal} {\bibinfo  {journal} {Appl. Phys. Lett.}\ }\textbf {\bibinfo
  {volume} {92}},\ \bibinfo {pages} {231909} (\bibinfo {year}
  {2008})}\BibitemShut {NoStop}%
\bibitem [{\citenamefont {Shreter}\ \emph {et~al.}(1997)\citenamefont
  {Shreter}, \citenamefont {Rebane}, \citenamefont {Davis}, \citenamefont
  {Barnard}, \citenamefont {Darbyshire}, \citenamefont {Steeds}, \citenamefont
  {Perry}, \citenamefont {Bremser},\ and\ \citenamefont
  {Davis}}]{Shreter_1997}%
  \BibitemOpen
  \bibfield  {author} {\bibinfo {author} {\bibfnamefont {Y.~G.}\ \bibnamefont
  {Shreter}}, \bibinfo {author} {\bibfnamefont {Y.~T.}\ \bibnamefont {Rebane}},
  \bibinfo {author} {\bibfnamefont {T.~J.}\ \bibnamefont {Davis}}, \bibinfo
  {author} {\bibfnamefont {J.}~\bibnamefont {Barnard}}, \bibinfo {author}
  {\bibfnamefont {M.}~\bibnamefont {Darbyshire}}, \bibinfo {author}
  {\bibfnamefont {J.~W.}\ \bibnamefont {Steeds}}, \bibinfo {author}
  {\bibfnamefont {W.~D.}\ \bibnamefont {Perry}}, \bibinfo {author}
  {\bibfnamefont {M.~D.}\ \bibnamefont {Bremser}},\ and\ \bibinfo {author}
  {\bibfnamefont {R.~F.}\ \bibnamefont {Davis}},\ }\bibfield  {title} {\bibinfo
  {title} {Disclocation luminescence in wurtzite {{GaN}}},\ }in\ \href
  {https://doi.org/10.1557/PROC-449-683} {\emph {\bibinfo {booktitle} {Mater.
  {{Res}}. {{Soc}}. {{Symp}}. {{Proc}}.}}},\ Vol.\ \bibinfo {volume} {449}\
  (\bibinfo {year} {1997})\ p.\ \bibinfo {pages} {683}\BibitemShut {NoStop}%
\bibitem [{\citenamefont {Reshchikov}\ \emph {et~al.}(2005)\citenamefont
  {Reshchikov}, \citenamefont {Huang}, \citenamefont {He}, \citenamefont
  {Morko{\c c}}, \citenamefont {Jasinski}, \citenamefont {{Liliental-Weber}},
  \citenamefont {Park},\ and\ \citenamefont {Lee}}]{Reshchikov_2005}%
  \BibitemOpen
  \bibfield  {author} {\bibinfo {author} {\bibfnamefont {M.}~\bibnamefont
  {Reshchikov}}, \bibinfo {author} {\bibfnamefont {D.}~\bibnamefont {Huang}},
  \bibinfo {author} {\bibfnamefont {L.}~\bibnamefont {He}}, \bibinfo {author}
  {\bibfnamefont {H.}~\bibnamefont {Morko{\c c}}}, \bibinfo {author}
  {\bibfnamefont {J.}~\bibnamefont {Jasinski}}, \bibinfo {author}
  {\bibfnamefont {Z.}~\bibnamefont {{Liliental-Weber}}}, \bibinfo {author}
  {\bibfnamefont {S.}~\bibnamefont {Park}},\ and\ \bibinfo {author}
  {\bibfnamefont {K.}~\bibnamefont {Lee}},\ }\bibfield  {title} {\bibinfo
  {title} {Manifestation of edge dislocations in photoluminescence of
  {{GaN}}},\ }\href {https://doi.org/10.1016/j.physb.2005.05.044} {\bibfield
  {journal} {\bibinfo  {journal} {Phys. B Condens. Matter}\ }\textbf {\bibinfo
  {volume} {367}},\ \bibinfo {pages} {35} (\bibinfo {year} {2005})}\BibitemShut
  {NoStop}%
\bibitem [{\citenamefont {Arslan}\ and\ \citenamefont
  {Browning}(2002)}]{Arslan_2002}%
  \BibitemOpen
  \bibfield  {author} {\bibinfo {author} {\bibfnamefont {I.}~\bibnamefont
  {Arslan}}\ and\ \bibinfo {author} {\bibfnamefont {N.~D.}\ \bibnamefont
  {Browning}},\ }\bibfield  {title} {\bibinfo {title} {Intrinsic electronic
  structure of threading dislocations in {{GaN}}},\ }\href
  {https://doi.org/10.1103/PhysRevB.65.075310} {\bibfield  {journal} {\bibinfo
  {journal} {Phys. Rev. B}\ }\textbf {\bibinfo {volume} {65}},\ \bibinfo
  {pages} {075310} (\bibinfo {year} {2002})}\BibitemShut {NoStop}%
\bibitem [{\citenamefont {Belabbas}\ \emph {et~al.}(2007)\citenamefont
  {Belabbas}, \citenamefont {B{\'e}r{\'e}}, \citenamefont {Chen}, \citenamefont
  {Petit}, \citenamefont {Belkhir}, \citenamefont {Ruterana},\ and\
  \citenamefont {Nouet}}]{Belabbas_2007}%
  \BibitemOpen
  \bibfield  {author} {\bibinfo {author} {\bibfnamefont {I.}~\bibnamefont
  {Belabbas}}, \bibinfo {author} {\bibfnamefont {A.}~\bibnamefont
  {B{\'e}r{\'e}}}, \bibinfo {author} {\bibfnamefont {J.}~\bibnamefont {Chen}},
  \bibinfo {author} {\bibfnamefont {S.}~\bibnamefont {Petit}}, \bibinfo
  {author} {\bibfnamefont {M.~A.}\ \bibnamefont {Belkhir}}, \bibinfo {author}
  {\bibfnamefont {P.}~\bibnamefont {Ruterana}},\ and\ \bibinfo {author}
  {\bibfnamefont {G.}~\bibnamefont {Nouet}},\ }\bibfield  {title} {\bibinfo
  {title} {Atomistic modeling of the (a+c)-mixed dislocation core in wurtzite
  {{GaN}}},\ }\href {https://doi.org/10.1103/PhysRevB.75.115201} {\bibfield
  {journal} {\bibinfo  {journal} {Phys. Rev. B}\ }\textbf {\bibinfo {volume}
  {75}},\ \bibinfo {pages} {115201} (\bibinfo {year} {2007})}\BibitemShut
  {NoStop}%
\bibitem [{\citenamefont {Matsubara}\ \emph {et~al.}(2013)\citenamefont
  {Matsubara}, \citenamefont {Godet}, \citenamefont {Pizzagalli},\ and\
  \citenamefont {Bellotti}}]{Matsubara_2013}%
  \BibitemOpen
  \bibfield  {author} {\bibinfo {author} {\bibfnamefont {M.}~\bibnamefont
  {Matsubara}}, \bibinfo {author} {\bibfnamefont {J.}~\bibnamefont {Godet}},
  \bibinfo {author} {\bibfnamefont {L.}~\bibnamefont {Pizzagalli}},\ and\
  \bibinfo {author} {\bibfnamefont {E.}~\bibnamefont {Bellotti}},\ }\bibfield
  {title} {\bibinfo {title} {Properties of threading screw dislocation core in
  wurtzite {{GaN}} studied by {{Heyd}}-{{Scuseria}}-{{Ernzerhof}} hybrid
  functional},\ }\href {https://doi.org/10.1063/1.4858618} {\bibfield
  {journal} {\bibinfo  {journal} {Appl. Phys. Lett.}\ }\textbf {\bibinfo
  {volume} {103}},\ \bibinfo {pages} {262107} (\bibinfo {year}
  {2013})}\BibitemShut {NoStop}%
\bibitem [{\citenamefont {Cottrell}\ and\ \citenamefont
  {Bilby}(1949)}]{Cottrell_1949}%
  \BibitemOpen
  \bibfield  {author} {\bibinfo {author} {\bibfnamefont {A.~H.}\ \bibnamefont
  {Cottrell}}\ and\ \bibinfo {author} {\bibfnamefont {B.~A.}\ \bibnamefont
  {Bilby}},\ }\bibfield  {title} {\bibinfo {title} {Dislocation theory of
  yielding and strain ageing of iron},\ }\href
  {https://doi.org/10.1088/0370-1298/62/1/308} {\bibfield  {journal} {\bibinfo
  {journal} {Proc. Phys. Soc. A}\ }\textbf {\bibinfo {volume} {62}},\ \bibinfo
  {pages} {49} (\bibinfo {year} {1949})}\BibitemShut {NoStop}%
\bibitem [{\citenamefont {Elsner}\ \emph {et~al.}(1998)\citenamefont {Elsner},
  \citenamefont {Jones}, \citenamefont {Heggie}, \citenamefont {Sitch},
  \citenamefont {Haugk}, \citenamefont {Frauenheim}, \citenamefont
  {{\"O}berg},\ and\ \citenamefont {Briddon}}]{Elsner_1998}%
  \BibitemOpen
  \bibfield  {author} {\bibinfo {author} {\bibfnamefont {J.}~\bibnamefont
  {Elsner}}, \bibinfo {author} {\bibfnamefont {R.}~\bibnamefont {Jones}},
  \bibinfo {author} {\bibfnamefont {M.~I.}\ \bibnamefont {Heggie}}, \bibinfo
  {author} {\bibfnamefont {P.~K.}\ \bibnamefont {Sitch}}, \bibinfo {author}
  {\bibfnamefont {M.}~\bibnamefont {Haugk}}, \bibinfo {author} {\bibfnamefont
  {T.}~\bibnamefont {Frauenheim}}, \bibinfo {author} {\bibfnamefont
  {S.}~\bibnamefont {{\"O}berg}},\ and\ \bibinfo {author} {\bibfnamefont
  {P.~R.}\ \bibnamefont {Briddon}},\ }\bibfield  {title} {\bibinfo {title}
  {Deep acceptors trapped at threading-edge dislocations in {{GaN}}},\ }\href
  {https://doi.org/10.1103/PhysRevB.58.12571} {\bibfield  {journal} {\bibinfo
  {journal} {Phys. Rev. B}\ }\textbf {\bibinfo {volume} {58}},\ \bibinfo
  {pages} {12571} (\bibinfo {year} {1998})}\BibitemShut {NoStop}%
\bibitem [{\citenamefont {Blumenau}\ \emph {et~al.}(2000)\citenamefont
  {Blumenau}, \citenamefont {Elsner}, \citenamefont {Jones}, \citenamefont
  {Heggie}, \citenamefont {{\"O}berg}, \citenamefont {Frauenheim},\ and\
  \citenamefont {Briddon}}]{Blumenau_2000}%
  \BibitemOpen
  \bibfield  {author} {\bibinfo {author} {\bibfnamefont {A.~T.}\ \bibnamefont
  {Blumenau}}, \bibinfo {author} {\bibfnamefont {J.}~\bibnamefont {Elsner}},
  \bibinfo {author} {\bibfnamefont {R.}~\bibnamefont {Jones}}, \bibinfo
  {author} {\bibfnamefont {M.~I.}\ \bibnamefont {Heggie}}, \bibinfo {author}
  {\bibfnamefont {S.}~\bibnamefont {{\"O}berg}}, \bibinfo {author}
  {\bibfnamefont {T.}~\bibnamefont {Frauenheim}},\ and\ \bibinfo {author}
  {\bibfnamefont {P.~R.}\ \bibnamefont {Briddon}},\ }\bibfield  {title}
  {\bibinfo {title} {Dislocations in hexagonal and cubic {{GaN}}},\ }\href
  {https://doi.org/10.1088/0953-8984/12/49/322} {\bibfield  {journal} {\bibinfo
   {journal} {J. Phys.: Condens. Matter}\ }\textbf {\bibinfo {volume} {12}},\
  \bibinfo {pages} {10223} (\bibinfo {year} {2000})}\BibitemShut {NoStop}%
\bibitem [{\citenamefont {Lee}\ \emph {et~al.}(2000)\citenamefont {Lee},
  \citenamefont {Belkhir}, \citenamefont {Zhu}, \citenamefont {Lee},
  \citenamefont {Hwang},\ and\ \citenamefont {Frauenheim}}]{Lee_2000}%
  \BibitemOpen
  \bibfield  {author} {\bibinfo {author} {\bibfnamefont {S.~M.}\ \bibnamefont
  {Lee}}, \bibinfo {author} {\bibfnamefont {M.~A.}\ \bibnamefont {Belkhir}},
  \bibinfo {author} {\bibfnamefont {X.~Y.}\ \bibnamefont {Zhu}}, \bibinfo
  {author} {\bibfnamefont {Y.~H.}\ \bibnamefont {Lee}}, \bibinfo {author}
  {\bibfnamefont {Y.~G.}\ \bibnamefont {Hwang}},\ and\ \bibinfo {author}
  {\bibfnamefont {T.}~\bibnamefont {Frauenheim}},\ }\bibfield  {title}
  {\bibinfo {title} {Electronic structures of {{GaN}} edge dislocations},\
  }\href {https://doi.org/10.1103/PhysRevB.61.16033} {\bibfield  {journal}
  {\bibinfo  {journal} {Phys. Rev. B}\ }\textbf {\bibinfo {volume} {61}},\
  \bibinfo {pages} {16033} (\bibinfo {year} {2000})}\BibitemShut {NoStop}%
\bibitem [{\citenamefont {Hirth}\ and\ \citenamefont
  {Lothe}(1982)}]{Hirth_1982}%
  \BibitemOpen
  \bibfield  {author} {\bibinfo {author} {\bibfnamefont {J.~P.}\ \bibnamefont
  {Hirth}}\ and\ \bibinfo {author} {\bibfnamefont {J.}~\bibnamefont {Lothe}},\
  }\href@noop {} {\emph {\bibinfo {title} {Theory of Dislocations}}},\ \bibinfo
  {edition} {second edition}\ ed.\ (\bibinfo  {publisher} {{Wiley}},\ \bibinfo
  {address} {{New York}},\ \bibinfo {year} {1982})\BibitemShut {NoStop}%
\bibitem [{\citenamefont {Gmeinwieser}\ \emph {et~al.}(2005)\citenamefont
  {Gmeinwieser}, \citenamefont {Gottfriedsen}, \citenamefont {Schwarz},
  \citenamefont {Wegscheider}, \citenamefont {Clos}, \citenamefont {Krtschil},
  \citenamefont {Krost}, \citenamefont {Weimar}, \citenamefont {Br{\"u}derl},
  \citenamefont {Lell},\ and\ \citenamefont {H{\"a}rle}}]{Gmeinwieser_2005}%
  \BibitemOpen
  \bibfield  {author} {\bibinfo {author} {\bibfnamefont {N.}~\bibnamefont
  {Gmeinwieser}}, \bibinfo {author} {\bibfnamefont {P.}~\bibnamefont
  {Gottfriedsen}}, \bibinfo {author} {\bibfnamefont {U.~T.}\ \bibnamefont
  {Schwarz}}, \bibinfo {author} {\bibfnamefont {W.}~\bibnamefont
  {Wegscheider}}, \bibinfo {author} {\bibfnamefont {R.}~\bibnamefont {Clos}},
  \bibinfo {author} {\bibfnamefont {A.}~\bibnamefont {Krtschil}}, \bibinfo
  {author} {\bibfnamefont {A.}~\bibnamefont {Krost}}, \bibinfo {author}
  {\bibfnamefont {A.}~\bibnamefont {Weimar}}, \bibinfo {author} {\bibfnamefont
  {G.}~\bibnamefont {Br{\"u}derl}}, \bibinfo {author} {\bibfnamefont
  {A.}~\bibnamefont {Lell}},\ and\ \bibinfo {author} {\bibfnamefont
  {V.}~\bibnamefont {H{\"a}rle}},\ }\bibfield  {title} {\bibinfo {title} {Local
  strain and potential distribution induced by single dislocations in
  {{GaN}}},\ }\href {https://doi.org/10.1063/1.2137879} {\bibfield  {journal}
  {\bibinfo  {journal} {J. Appl. Phys.}\ }\textbf {\bibinfo {volume} {98}},\
  \bibinfo {pages} {116102} (\bibinfo {year} {2005})}\BibitemShut {NoStop}%
\bibitem [{\citenamefont {Gmeinwieser}\ and\ \citenamefont
  {Schwarz}(2007)}]{Gmeinwieser_2007}%
  \BibitemOpen
  \bibfield  {author} {\bibinfo {author} {\bibfnamefont {N.}~\bibnamefont
  {Gmeinwieser}}\ and\ \bibinfo {author} {\bibfnamefont {U.~T.}\ \bibnamefont
  {Schwarz}},\ }\bibfield  {title} {\bibinfo {title} {Pattern formation and
  directional and spatial ordering of edge dislocations in bulk {{GaN}}:
  Microphotoluminescence spectra and continuum elastic calculations},\ }\href
  {https://doi.org/10.1103/PhysRevB.75.245213} {\bibfield  {journal} {\bibinfo
  {journal} {Phys. Rev. B}\ }\textbf {\bibinfo {volume} {75}},\ \bibinfo
  {pages} {245213} (\bibinfo {year} {2007})}\BibitemShut {NoStop}%
\bibitem [{\citenamefont {Liu}\ \emph {et~al.}(2016)\citenamefont {Liu},
  \citenamefont {Carlin}, \citenamefont {Grandjean}, \citenamefont {Deveaud},\
  and\ \citenamefont {Jacopin}}]{Liu_2016}%
  \BibitemOpen
  \bibfield  {author} {\bibinfo {author} {\bibfnamefont {W.}~\bibnamefont
  {Liu}}, \bibinfo {author} {\bibfnamefont {J.-F.}\ \bibnamefont {Carlin}},
  \bibinfo {author} {\bibfnamefont {N.}~\bibnamefont {Grandjean}}, \bibinfo
  {author} {\bibfnamefont {B.}~\bibnamefont {Deveaud}},\ and\ \bibinfo {author}
  {\bibfnamefont {G.}~\bibnamefont {Jacopin}},\ }\bibfield  {title} {\bibinfo
  {title} {Exciton dynamics at a single dislocation in {{GaN}} probed by
  picosecond time-resolved cathodoluminescence},\ }\href
  {https://doi.org/10.1063/1.4959832} {\bibfield  {journal} {\bibinfo
  {journal} {Appl. Phys. Lett.}\ }\textbf {\bibinfo {volume} {109}},\ \bibinfo
  {pages} {042101} (\bibinfo {year} {2016})}\BibitemShut {NoStop}%
\bibitem [{\citenamefont {Smirnova}(1974)}]{Smirnova_1974}%
  \BibitemOpen
  \bibfield  {author} {\bibinfo {author} {\bibfnamefont {I.~S.}\ \bibnamefont
  {Smirnova}},\ }\bibfield  {title} {\bibinfo {title} {Electric fields around
  dislocations in crystals having the wurtzite structure},\ }\href@noop {}
  {\bibfield  {journal} {\bibinfo  {journal} {Sov. Phys. Solid State}\ }\textbf
  {\bibinfo {volume} {15}},\ \bibinfo {pages} {1543} (\bibinfo {year}
  {1974})}\BibitemShut {NoStop}%
\bibitem [{\citenamefont {Shi}\ \emph {et~al.}(1999)\citenamefont {Shi},
  \citenamefont {Asbeck},\ and\ \citenamefont {Yu}}]{Shi_1999}%
  \BibitemOpen
  \bibfield  {author} {\bibinfo {author} {\bibfnamefont {C.}~\bibnamefont
  {Shi}}, \bibinfo {author} {\bibfnamefont {P.~M.}\ \bibnamefont {Asbeck}},\
  and\ \bibinfo {author} {\bibfnamefont {E.~T.}\ \bibnamefont {Yu}},\
  }\bibfield  {title} {\bibinfo {title} {Piezoelectric polarization associated
  with dislocations in wurtzite {{GaN}}},\ }\href
  {https://doi.org/10.1063/1.123149} {\bibfield  {journal} {\bibinfo  {journal}
  {Appl. Phys. Lett.}\ }\textbf {\bibinfo {volume} {74}},\ \bibinfo {pages}
  {573} (\bibinfo {year} {1999})}\BibitemShut {NoStop}%
\bibitem [{\citenamefont {Yoffe}(1961)}]{Yoffe_1961}%
  \BibitemOpen
  \bibfield  {author} {\bibinfo {author} {\bibfnamefont {E.~H.}\ \bibnamefont
  {Yoffe}},\ }\bibfield  {title} {\bibinfo {title} {A dislocation at a free
  surface},\ }\href {https://doi.org/10.1080/14786436108239675} {\bibfield
  {journal} {\bibinfo  {journal} {Philos. Mag.}\ }\textbf {\bibinfo {volume}
  {6}},\ \bibinfo {pages} {1147} (\bibinfo {year} {1961})}\BibitemShut
  {NoStop}%
\bibitem [{\citenamefont {Taupin}\ \emph {et~al.}(2014)\citenamefont {Taupin},
  \citenamefont {Fressengeas}, \citenamefont {Ventura}, \citenamefont
  {Lebyodkin},\ and\ \citenamefont {Gornakov}}]{Taupin_2014}%
  \BibitemOpen
  \bibfield  {author} {\bibinfo {author} {\bibfnamefont {V.}~\bibnamefont
  {Taupin}}, \bibinfo {author} {\bibfnamefont {C.}~\bibnamefont {Fressengeas}},
  \bibinfo {author} {\bibfnamefont {P.}~\bibnamefont {Ventura}}, \bibinfo
  {author} {\bibfnamefont {M.}~\bibnamefont {Lebyodkin}},\ and\ \bibinfo
  {author} {\bibfnamefont {V.}~\bibnamefont {Gornakov}},\ }\bibfield  {title}
  {\bibinfo {title} {A field theory of piezoelectric media containing
  dislocations},\ }\href {https://doi.org/10.1063/1.4870931} {\bibfield
  {journal} {\bibinfo  {journal} {J. Appl. Phys.}\ }\textbf {\bibinfo {volume}
  {115}},\ \bibinfo {pages} {144902} (\bibinfo {year} {2014})}\BibitemShut
  {NoStop}%
\bibitem [{\citenamefont {Kaganer}\ \emph {et~al.}(2018)\citenamefont
  {Kaganer}, \citenamefont {Sabelfeld},\ and\ \citenamefont
  {Brandt}}]{Kaganer_2018}%
  \BibitemOpen
  \bibfield  {author} {\bibinfo {author} {\bibfnamefont {V.~M.}\ \bibnamefont
  {Kaganer}}, \bibinfo {author} {\bibfnamefont {K.~K.}\ \bibnamefont
  {Sabelfeld}},\ and\ \bibinfo {author} {\bibfnamefont {O.}~\bibnamefont
  {Brandt}},\ }\bibfield  {title} {\bibinfo {title} {Piezoelectric field,
  exciton lifetime, and cathodoluminescence intensity at threading dislocations
  in {{GaN}}\{0001\}},\ }\href {https://doi.org/10.1063/1.5022170} {\bibfield
  {journal} {\bibinfo  {journal} {Appl. Phys. Lett.}\ }\textbf {\bibinfo
  {volume} {112}},\ \bibinfo {pages} {122101} (\bibinfo {year}
  {2018})}\BibitemShut {NoStop}%
\bibitem [{\citenamefont {Kaganer}\ \emph {et~al.}(2019)\citenamefont
  {Kaganer}, \citenamefont {L{\"a}hnemann}, \citenamefont {Pf{\"u}ller},
  \citenamefont {Sabelfeld}, \citenamefont {Kireeva},\ and\ \citenamefont
  {Brandt}}]{Kaganer_2019}%
  \BibitemOpen
  \bibfield  {author} {\bibinfo {author} {\bibfnamefont {V.~M.}\ \bibnamefont
  {Kaganer}}, \bibinfo {author} {\bibfnamefont {J.}~\bibnamefont
  {L{\"a}hnemann}}, \bibinfo {author} {\bibfnamefont {C.}~\bibnamefont
  {Pf{\"u}ller}}, \bibinfo {author} {\bibfnamefont {K.~K.}\ \bibnamefont
  {Sabelfeld}}, \bibinfo {author} {\bibfnamefont {A.~E.}\ \bibnamefont
  {Kireeva}},\ and\ \bibinfo {author} {\bibfnamefont {O.}~\bibnamefont
  {Brandt}},\ }\bibfield  {title} {\bibinfo {title} {Determination of the
  carrier diffusion length in {{GaN}} from cathodoluminescence maps around
  threading dislocations: Fallacies and opportunities},\ }\href
  {https://doi.org/10.1103/PhysRevApplied.12.054038} {\bibfield  {journal}
  {\bibinfo  {journal} {Phys. Rev. Appl.}\ }\textbf {\bibinfo {volume} {12}},\
  \bibinfo {pages} {054038} (\bibinfo {year} {2019})}\BibitemShut {NoStop}%
\bibitem [{\citenamefont {Cherns}\ \emph {et~al.}(2001)\citenamefont {Cherns},
  \citenamefont {Henley},\ and\ \citenamefont {Ponce}}]{Cherns_2001}%
  \BibitemOpen
  \bibfield  {author} {\bibinfo {author} {\bibfnamefont {D.}~\bibnamefont
  {Cherns}}, \bibinfo {author} {\bibfnamefont {S.~J.}\ \bibnamefont {Henley}},\
  and\ \bibinfo {author} {\bibfnamefont {F.~A.}\ \bibnamefont {Ponce}},\
  }\bibfield  {title} {\bibinfo {title} {Edge and screw dislocations as
  nonradiative centers in {{InGaN}}/{{GaN}} quantum well luminescence},\ }\href
  {https://doi.org/10.1063/1.1369610} {\bibfield  {journal} {\bibinfo
  {journal} {Appl. Phys. Lett.}\ }\textbf {\bibinfo {volume} {78}},\ \bibinfo
  {pages} {2691} (\bibinfo {year} {2001})}\BibitemShut {NoStop}%
\bibitem [{\citenamefont {Miyajima}\ \emph {et~al.}(2001)\citenamefont
  {Miyajima}, \citenamefont {Hino}, \citenamefont {Tomiya}, \citenamefont
  {Yanashima}, \citenamefont {Nakajima}, \citenamefont {Nanishi}, \citenamefont
  {Satake}, \citenamefont {Masumoto}, \citenamefont {Akimoto}, \citenamefont
  {Kobayashi},\ and\ \citenamefont {Ikeda}}]{Miyajima_2001}%
  \BibitemOpen
  \bibfield  {author} {\bibinfo {author} {\bibfnamefont {T.}~\bibnamefont
  {Miyajima}}, \bibinfo {author} {\bibfnamefont {T.}~\bibnamefont {Hino}},
  \bibinfo {author} {\bibfnamefont {S.}~\bibnamefont {Tomiya}}, \bibinfo
  {author} {\bibfnamefont {K.}~\bibnamefont {Yanashima}}, \bibinfo {author}
  {\bibfnamefont {H.}~\bibnamefont {Nakajima}}, \bibinfo {author}
  {\bibfnamefont {Y.}~\bibnamefont {Nanishi}}, \bibinfo {author} {\bibfnamefont
  {A.}~\bibnamefont {Satake}}, \bibinfo {author} {\bibfnamefont
  {Y.}~\bibnamefont {Masumoto}}, \bibinfo {author} {\bibfnamefont
  {K.}~\bibnamefont {Akimoto}}, \bibinfo {author} {\bibfnamefont
  {T.}~\bibnamefont {Kobayashi}},\ and\ \bibinfo {author} {\bibfnamefont
  {M.}~\bibnamefont {Ikeda}},\ }\bibfield  {title} {\bibinfo {title} {Threading
  dislocations and optical properties of {{GaN}} and {{GaInN}}},\ }\href
  {https://doi.org/10.1002/1521-3951(200111)228:2<395::AID-PSSB395>3.0.CO;2-2}
  {\bibfield  {journal} {\bibinfo  {journal} {Phys. Status Solidi B}\ }\textbf
  {\bibinfo {volume} {228}},\ \bibinfo {pages} {395} (\bibinfo {year}
  {2001})}\BibitemShut {NoStop}%
\bibitem [{\citenamefont {Remmele}\ \emph {et~al.}(2001)\citenamefont
  {Remmele}, \citenamefont {Albrecht}, \citenamefont {Strunk}, \citenamefont
  {Blumenau}, \citenamefont {Heggie}, \citenamefont {Elsner}, \citenamefont
  {Frauenheim}, \citenamefont {Schenk},\ and\ \citenamefont
  {Gibart}}]{Remmele_2001}%
  \BibitemOpen
  \bibfield  {author} {\bibinfo {author} {\bibfnamefont {T.}~\bibnamefont
  {Remmele}}, \bibinfo {author} {\bibfnamefont {M.}~\bibnamefont {Albrecht}},
  \bibinfo {author} {\bibfnamefont {H.~P.}\ \bibnamefont {Strunk}}, \bibinfo
  {author} {\bibfnamefont {A.~T.}\ \bibnamefont {Blumenau}}, \bibinfo {author}
  {\bibfnamefont {M.~I.}\ \bibnamefont {Heggie}}, \bibinfo {author}
  {\bibfnamefont {J.}~\bibnamefont {Elsner}}, \bibinfo {author} {\bibfnamefont
  {T.}~\bibnamefont {Frauenheim}}, \bibinfo {author} {\bibfnamefont {H.~P.~D.}\
  \bibnamefont {Schenk}},\ and\ \bibinfo {author} {\bibfnamefont
  {P.}~\bibnamefont {Gibart}},\ }\bibfield  {title} {\bibinfo {title} {Core
  structure of dislocations in {{GaN}} revealed by transmission electron
  microscopy},\ }in\ \href {https://doi.org/10.1201/9781351074629} {\emph
  {\bibinfo {booktitle} {Inst. {{Phys}}. {{Conf}}. {{Ser}}.}}},\ Vol.\ \bibinfo
  {volume} {169}\ (\bibinfo {year} {2001})\ pp.\ \bibinfo {pages}
  {323--326}\BibitemShut {NoStop}%
\bibitem [{\citenamefont {Yamamoto}\ \emph {et~al.}(2003)\citenamefont
  {Yamamoto}, \citenamefont {Itoh}, \citenamefont {Grillo}, \citenamefont
  {Chichibu}, \citenamefont {Keller}, \citenamefont {Speck}, \citenamefont
  {DenBaars}, \citenamefont {Mishra}, \citenamefont {Nakamura},\ and\
  \citenamefont {Salviati}}]{Yamamoto_2003}%
  \BibitemOpen
  \bibfield  {author} {\bibinfo {author} {\bibfnamefont {N.}~\bibnamefont
  {Yamamoto}}, \bibinfo {author} {\bibfnamefont {H.}~\bibnamefont {Itoh}},
  \bibinfo {author} {\bibfnamefont {V.}~\bibnamefont {Grillo}}, \bibinfo
  {author} {\bibfnamefont {S.~F.}\ \bibnamefont {Chichibu}}, \bibinfo {author}
  {\bibfnamefont {S.}~\bibnamefont {Keller}}, \bibinfo {author} {\bibfnamefont
  {J.~S.}\ \bibnamefont {Speck}}, \bibinfo {author} {\bibfnamefont {S.~P.}\
  \bibnamefont {DenBaars}}, \bibinfo {author} {\bibfnamefont {U.~K.}\
  \bibnamefont {Mishra}}, \bibinfo {author} {\bibfnamefont {S.}~\bibnamefont
  {Nakamura}},\ and\ \bibinfo {author} {\bibfnamefont {G.}~\bibnamefont
  {Salviati}},\ }\bibfield  {title} {\bibinfo {title} {Cathodoluminescence
  characterization of dislocations in gallium nitride using a transmission
  electron microscope},\ }\href {https://doi.org/10.1063/1.1598632} {\bibfield
  {journal} {\bibinfo  {journal} {J. Appl. Phys.}\ }\textbf {\bibinfo {volume}
  {94}},\ \bibinfo {pages} {4315} (\bibinfo {year} {2003})}\BibitemShut
  {NoStop}%
\bibitem [{\citenamefont {{Naresh-Kumar}}\ \emph {et~al.}(2014)\citenamefont
  {{Naresh-Kumar}}, \citenamefont {Bruckbauer}, \citenamefont {Edwards},
  \citenamefont {Kraeusel}, \citenamefont {Hourahine}, \citenamefont {Martin},
  \citenamefont {Kappers}, \citenamefont {Moram}, \citenamefont {Lovelock},
  \citenamefont {Oliver}, \citenamefont {Humphreys},\ and\ \citenamefont
  {{Trager-Cowan}}}]{Naresh-Kumar_2014}%
  \BibitemOpen
  \bibfield  {author} {\bibinfo {author} {\bibfnamefont {G.}~\bibnamefont
  {{Naresh-Kumar}}}, \bibinfo {author} {\bibfnamefont {J.}~\bibnamefont
  {Bruckbauer}}, \bibinfo {author} {\bibfnamefont {P.~R.}\ \bibnamefont
  {Edwards}}, \bibinfo {author} {\bibfnamefont {S.}~\bibnamefont {Kraeusel}},
  \bibinfo {author} {\bibfnamefont {B.}~\bibnamefont {Hourahine}}, \bibinfo
  {author} {\bibfnamefont {R.~W.}\ \bibnamefont {Martin}}, \bibinfo {author}
  {\bibfnamefont {M.~J.}\ \bibnamefont {Kappers}}, \bibinfo {author}
  {\bibfnamefont {M.~A.}\ \bibnamefont {Moram}}, \bibinfo {author}
  {\bibfnamefont {S.}~\bibnamefont {Lovelock}}, \bibinfo {author}
  {\bibfnamefont {R.~A.}\ \bibnamefont {Oliver}}, \bibinfo {author}
  {\bibfnamefont {C.~J.}\ \bibnamefont {Humphreys}},\ and\ \bibinfo {author}
  {\bibfnamefont {C.}~\bibnamefont {{Trager-Cowan}}},\ }\bibfield  {title}
  {\bibinfo {title} {Coincident electron channeling and cathodoluminescence
  studies of threading dislocations in {{GaN}}},\ }\href
  {https://doi.org/10.1017/S1431927613013755} {\bibfield  {journal} {\bibinfo
  {journal} {Microsc. Microanal.}\ }\textbf {\bibinfo {volume} {20}},\ \bibinfo
  {pages} {55} (\bibinfo {year} {2014})}\BibitemShut {NoStop}%
\bibitem [{\citenamefont {Hurni}\ \emph {et~al.}(2015)\citenamefont {Hurni},
  \citenamefont {David}, \citenamefont {Cich}, \citenamefont {Aldaz},
  \citenamefont {Ellis}, \citenamefont {Huang}, \citenamefont {Tyagi},
  \citenamefont {DeLille}, \citenamefont {Craven}, \citenamefont {Steranka},\
  and\ \citenamefont {Krames}}]{Hurni_2015}%
  \BibitemOpen
  \bibfield  {author} {\bibinfo {author} {\bibfnamefont {C.~A.}\ \bibnamefont
  {Hurni}}, \bibinfo {author} {\bibfnamefont {A.}~\bibnamefont {David}},
  \bibinfo {author} {\bibfnamefont {M.~J.}\ \bibnamefont {Cich}}, \bibinfo
  {author} {\bibfnamefont {R.~I.}\ \bibnamefont {Aldaz}}, \bibinfo {author}
  {\bibfnamefont {B.}~\bibnamefont {Ellis}}, \bibinfo {author} {\bibfnamefont
  {K.}~\bibnamefont {Huang}}, \bibinfo {author} {\bibfnamefont
  {A.}~\bibnamefont {Tyagi}}, \bibinfo {author} {\bibfnamefont {R.~A.}\
  \bibnamefont {DeLille}}, \bibinfo {author} {\bibfnamefont {M.~D.}\
  \bibnamefont {Craven}}, \bibinfo {author} {\bibfnamefont {F.~M.}\
  \bibnamefont {Steranka}},\ and\ \bibinfo {author} {\bibfnamefont {M.~R.}\
  \bibnamefont {Krames}},\ }\bibfield  {title} {\bibinfo {title} {Bulk {{GaN}}
  flip-chip violet light-emitting diodes with optimized efficiency for
  high-power operation},\ }\href {https://doi.org/10.1063/1.4905873} {\bibfield
   {journal} {\bibinfo  {journal} {Appl. Phys. Lett.}\ }\textbf {\bibinfo
  {volume} {106}},\ \bibinfo {pages} {031101} (\bibinfo {year}
  {2015})}\BibitemShut {NoStop}%
\bibitem [{\citenamefont {Jahn}\ \emph {et~al.}(2020)\citenamefont {Jahn},
  \citenamefont {Kaganer}, \citenamefont {Sabelfeld}, \citenamefont {Kireeva},
  \citenamefont {L{\"a}hnemann}, \citenamefont {Pf{\"u}ller}, \citenamefont
  {Ch{\`e}ze}, \citenamefont {Biermann}, \citenamefont {Calarco},\ and\
  \citenamefont {Brandt}}]{Jahn_2020}%
  \BibitemOpen
  \bibfield  {author} {\bibinfo {author} {\bibfnamefont {U.}~\bibnamefont
  {Jahn}}, \bibinfo {author} {\bibfnamefont {V.~M.}\ \bibnamefont {Kaganer}},
  \bibinfo {author} {\bibfnamefont {K.~K.}\ \bibnamefont {Sabelfeld}}, \bibinfo
  {author} {\bibfnamefont {A.~E.}\ \bibnamefont {Kireeva}}, \bibinfo {author}
  {\bibfnamefont {J.}~\bibnamefont {L{\"a}hnemann}}, \bibinfo {author}
  {\bibfnamefont {C.}~\bibnamefont {Pf{\"u}ller}}, \bibinfo {author}
  {\bibfnamefont {C.}~\bibnamefont {Ch{\`e}ze}}, \bibinfo {author}
  {\bibfnamefont {K.}~\bibnamefont {Biermann}}, \bibinfo {author}
  {\bibfnamefont {R.}~\bibnamefont {Calarco}},\ and\ \bibinfo {author}
  {\bibfnamefont {O.}~\bibnamefont {Brandt}},\ }\bibfield  {title} {\bibinfo
  {title} {Carrier diffusion in {{GaN}}---a cathodoluminescence study. {{I}}:
  Temperature-dependent generation volume},\ }\href
  {http://arxiv.org/abs/2002.08713} {\bibfield  {journal} {\bibinfo  {journal}
  {arXiv:2002.08713 [cond-mat.mtrl-sci, physics.app-ph]}\ } (\bibinfo {year}
  {2020})}\BibitemShut {NoStop}%
\bibitem [{\citenamefont {Brandt}\ \emph {et~al.}(2020)\citenamefont {Brandt},
  \citenamefont {Kaganer}, \citenamefont {L{\"a}hnemann}, \citenamefont
  {Flissikowski}, \citenamefont {Pf{\"u}ller}, \citenamefont {Sabelfeld},
  \citenamefont {Kireeva}, \citenamefont {Ch{\`e}ze}, \citenamefont {Calarco},
  \citenamefont {Grahn},\ and\ \citenamefont {Jahn}}]{Brandt_2020}%
  \BibitemOpen
  \bibfield  {author} {\bibinfo {author} {\bibfnamefont {O.}~\bibnamefont
  {Brandt}}, \bibinfo {author} {\bibfnamefont {V.~M.}\ \bibnamefont {Kaganer}},
  \bibinfo {author} {\bibfnamefont {J.}~\bibnamefont {L{\"a}hnemann}}, \bibinfo
  {author} {\bibfnamefont {T.}~\bibnamefont {Flissikowski}}, \bibinfo {author}
  {\bibfnamefont {C.}~\bibnamefont {Pf{\"u}ller}}, \bibinfo {author}
  {\bibfnamefont {K.~K.}\ \bibnamefont {Sabelfeld}}, \bibinfo {author}
  {\bibfnamefont {A.~E.}\ \bibnamefont {Kireeva}}, \bibinfo {author}
  {\bibfnamefont {C.}~\bibnamefont {Ch{\`e}ze}}, \bibinfo {author}
  {\bibfnamefont {R.}~\bibnamefont {Calarco}}, \bibinfo {author} {\bibfnamefont
  {H.~T.}\ \bibnamefont {Grahn}},\ and\ \bibinfo {author} {\bibfnamefont
  {U.}~\bibnamefont {Jahn}},\ }\bibfield  {title} {\bibinfo {title} {Carrier
  diffusion in {{GaN}} -- a cathodoluminescence study. {{II}}: Ambipolar vs.
  exciton diffusion},\ }\href {http://arxiv.org/abs/2009.13983} {\bibfield
  {journal} {\bibinfo  {journal} {arXiv:2009.13983 [cond-mat,
  physics:physics]}\ } (\bibinfo {year} {2020})}\BibitemShut {NoStop}%
\bibitem [{\citenamefont {Ebeling}\ \emph {et~al.}(1976)\citenamefont
  {Ebeling}, \citenamefont {Kraeft},\ and\ \citenamefont
  {Kremp}}]{Ebeling_1976}%
  \BibitemOpen
  \bibfield  {author} {\bibinfo {author} {\bibfnamefont {W.}~\bibnamefont
  {Ebeling}}, \bibinfo {author} {\bibfnamefont {W.~D.}\ \bibnamefont
  {Kraeft}},\ and\ \bibinfo {author} {\bibfnamefont {D.}~\bibnamefont
  {Kremp}},\ }\href@noop {} {\emph {\bibinfo {title} {Theory of Bound States
  and Ionization Equilibrium in Plasmas and Solids}}}\ (\bibinfo  {publisher}
  {{Akademie-Verlag}},\ \bibinfo {address} {{Berlin}},\ \bibinfo {year}
  {1976})\BibitemShut {NoStop}%
\bibitem [{\citenamefont {Brandt}\ \emph {et~al.}(1998)\citenamefont {Brandt},
  \citenamefont {Yang}, \citenamefont {W{\"u}nsche}, \citenamefont {Jahn},
  \citenamefont {Ringling}, \citenamefont {Paris}, \citenamefont {Grahn},\ and\
  \citenamefont {Ploog}}]{Brandt_1998}%
  \BibitemOpen
  \bibfield  {author} {\bibinfo {author} {\bibfnamefont {O.}~\bibnamefont
  {Brandt}}, \bibinfo {author} {\bibfnamefont {B.}~\bibnamefont {Yang}},
  \bibinfo {author} {\bibfnamefont {H.-J.}\ \bibnamefont {W{\"u}nsche}},
  \bibinfo {author} {\bibfnamefont {U.}~\bibnamefont {Jahn}}, \bibinfo {author}
  {\bibfnamefont {J.}~\bibnamefont {Ringling}}, \bibinfo {author}
  {\bibfnamefont {G.}~\bibnamefont {Paris}}, \bibinfo {author} {\bibfnamefont
  {H.~T.}\ \bibnamefont {Grahn}},\ and\ \bibinfo {author} {\bibfnamefont
  {K.~H.}\ \bibnamefont {Ploog}},\ }\bibfield  {title} {\bibinfo {title}
  {Impact of exciton diffusion on the optical properties of thin {{GaN}}
  layers},\ }\href {https://doi.org/10.1103/physrevb.58.r13407} {\bibfield
  {journal} {\bibinfo  {journal} {Phys. Rev. B}\ }\textbf {\bibinfo {volume}
  {58}},\ \bibinfo {pages} {R13407} (\bibinfo {year} {1998})}\BibitemShut
  {NoStop}%
\bibitem [{\citenamefont {Toth}\ and\ \citenamefont
  {Phillips}(1998)}]{Toth_1998}%
  \BibitemOpen
  \bibfield  {author} {\bibinfo {author} {\bibfnamefont {M.}~\bibnamefont
  {Toth}}\ and\ \bibinfo {author} {\bibfnamefont {M.~R.}\ \bibnamefont
  {Phillips}},\ }\bibfield  {title} {\bibinfo {title} {Monte {{Carlo}} modeling
  of cathodoluminescence generation using electron energy loss curves},\ }\href
  {https://doi.org/10.1002/sca.1998.4950200601} {\bibfield  {journal} {\bibinfo
   {journal} {Scanning}\ }\textbf {\bibinfo {volume} {20}},\ \bibinfo {pages}
  {425} (\bibinfo {year} {1998})}\BibitemShut {NoStop}%
\bibitem [{\citenamefont {Demers}\ \emph {et~al.}(2011)\citenamefont {Demers},
  \citenamefont {{Poirier-Demers}}, \citenamefont {Couture}, \citenamefont
  {Joly}, \citenamefont {Guilmain}, \citenamefont {{de Jonge}},\ and\
  \citenamefont {Drouin}}]{Demers_2011}%
  \BibitemOpen
  \bibfield  {author} {\bibinfo {author} {\bibfnamefont {H.}~\bibnamefont
  {Demers}}, \bibinfo {author} {\bibfnamefont {N.}~\bibnamefont
  {{Poirier-Demers}}}, \bibinfo {author} {\bibfnamefont {A.~R.}\ \bibnamefont
  {Couture}}, \bibinfo {author} {\bibfnamefont {D.}~\bibnamefont {Joly}},
  \bibinfo {author} {\bibfnamefont {M.}~\bibnamefont {Guilmain}}, \bibinfo
  {author} {\bibfnamefont {N.}~\bibnamefont {{de Jonge}}},\ and\ \bibinfo
  {author} {\bibfnamefont {D.}~\bibnamefont {Drouin}},\ }\bibfield  {title}
  {\bibinfo {title} {Three-dimensional electron microscopy simulation with the
  {{CASINO Monte Carlo}} software},\ }\href {https://doi.org/10.1002/sca.20262}
  {\bibfield  {journal} {\bibinfo  {journal} {Scanning}\ }\textbf {\bibinfo
  {volume} {33}},\ \bibinfo {pages} {135} (\bibinfo {year} {2011})}\BibitemShut
  {NoStop}%
\bibitem [{\citenamefont {{de la Pe{\~n}a}}\ \emph {et~al.}(2019)\citenamefont
  {{de la Pe{\~n}a}}, \citenamefont {Prestat}, \citenamefont {Fauske},
  \citenamefont {Burdet}, \citenamefont {Jokubauskas}, \citenamefont {Nord},
  \citenamefont {Ostasevicius}, \citenamefont {MacArthur}, \citenamefont
  {Sarahan}, \citenamefont {Johnstone}, \citenamefont {Taillon}, \citenamefont
  {L{\"a}hnemann}, \citenamefont {Migunov}, \citenamefont {Eljarrat},
  \citenamefont {Caron}, \citenamefont {Aarholt}, \citenamefont {Mazzucco},
  \citenamefont {Walls}, \citenamefont {Slater}, \citenamefont {Winkler},
  \citenamefont {{pquinn-dls}}, \citenamefont {Martineau}, \citenamefont
  {Donval}, \citenamefont {McLeod}, \citenamefont {Hoglund}, \citenamefont
  {Alxneit}, \citenamefont {Lundeby}, \citenamefont {Henninen}, \citenamefont
  {Zagonel},\ and\ \citenamefont {Garmannslund}}]{Hyperspy}%
  \BibitemOpen
  \bibfield  {author} {\bibinfo {author} {\bibfnamefont {F.}~\bibnamefont {{de
  la Pe{\~n}a}}}, \bibinfo {author} {\bibfnamefont {E.}~\bibnamefont
  {Prestat}}, \bibinfo {author} {\bibfnamefont {V.~T.}\ \bibnamefont {Fauske}},
  \bibinfo {author} {\bibfnamefont {P.}~\bibnamefont {Burdet}}, \bibinfo
  {author} {\bibfnamefont {P.}~\bibnamefont {Jokubauskas}}, \bibinfo {author}
  {\bibfnamefont {M.}~\bibnamefont {Nord}}, \bibinfo {author} {\bibfnamefont
  {T.}~\bibnamefont {Ostasevicius}}, \bibinfo {author} {\bibfnamefont {K.~E.}\
  \bibnamefont {MacArthur}}, \bibinfo {author} {\bibfnamefont {M.}~\bibnamefont
  {Sarahan}}, \bibinfo {author} {\bibfnamefont {D.~N.}\ \bibnamefont
  {Johnstone}}, \bibinfo {author} {\bibfnamefont {J.}~\bibnamefont {Taillon}},
  \bibinfo {author} {\bibfnamefont {J.}~\bibnamefont {L{\"a}hnemann}}, \bibinfo
  {author} {\bibfnamefont {V.}~\bibnamefont {Migunov}}, \bibinfo {author}
  {\bibfnamefont {A.}~\bibnamefont {Eljarrat}}, \bibinfo {author}
  {\bibfnamefont {J.}~\bibnamefont {Caron}}, \bibinfo {author} {\bibfnamefont
  {T.}~\bibnamefont {Aarholt}}, \bibinfo {author} {\bibfnamefont
  {S.}~\bibnamefont {Mazzucco}}, \bibinfo {author} {\bibfnamefont
  {M.}~\bibnamefont {Walls}}, \bibinfo {author} {\bibfnamefont
  {T.}~\bibnamefont {Slater}}, \bibinfo {author} {\bibfnamefont
  {F.}~\bibnamefont {Winkler}}, \bibinfo {author} {\bibnamefont
  {{pquinn-dls}}}, \bibinfo {author} {\bibfnamefont {B.}~\bibnamefont
  {Martineau}}, \bibinfo {author} {\bibfnamefont {G.}~\bibnamefont {Donval}},
  \bibinfo {author} {\bibfnamefont {R.}~\bibnamefont {McLeod}}, \bibinfo
  {author} {\bibfnamefont {E.~R.}\ \bibnamefont {Hoglund}}, \bibinfo {author}
  {\bibfnamefont {I.}~\bibnamefont {Alxneit}}, \bibinfo {author} {\bibfnamefont
  {D.}~\bibnamefont {Lundeby}}, \bibinfo {author} {\bibfnamefont
  {T.}~\bibnamefont {Henninen}}, \bibinfo {author} {\bibfnamefont {L.~F.}\
  \bibnamefont {Zagonel}},\ and\ \bibinfo {author} {\bibfnamefont
  {A.}~\bibnamefont {Garmannslund}},\ }\href
  {https://doi.org/10.5281/zenodo.3396791} {\bibinfo {title} {{{HyperSpy}}
  v1.5.2}} (\bibinfo {year} {2019})\BibitemShut {NoStop}%
\bibitem [{\citenamefont {Sabelfeld}(1991)}]{Sabelfeld_1991}%
  \BibitemOpen
  \bibfield  {author} {\bibinfo {author} {\bibfnamefont {K.~K.}\ \bibnamefont
  {Sabelfeld}},\ }\href {https://www.springer.com/gp/book/9783642759796} {\emph
  {\bibinfo {title} {Monte {{Carlo}} Methods in Boundary Value Problems}}}\
  (\bibinfo  {publisher} {{Springer-Verlag}},\ \bibinfo {address} {{Berlin
  Heidelberg}},\ \bibinfo {year} {1991})\BibitemShut {NoStop}%
\bibitem [{\citenamefont {Sabelfeld}(2016)}]{Sabelfeld_2016}%
  \BibitemOpen
  \bibfield  {author} {\bibinfo {author} {\bibfnamefont {K.~K.}\ \bibnamefont
  {Sabelfeld}},\ }\bibfield  {title} {\bibinfo {title} {Random walk on spheres
  method for solving drift-diffusion problems},\ }\href
  {https://doi.org/10.1515/mcma-2016-0118} {\bibfield  {journal} {\bibinfo
  {journal} {Monte Carlo Methods Appl.}\ }\textbf {\bibinfo {volume} {22}},\
  \bibinfo {pages} {265} (\bibinfo {year} {2016})}\BibitemShut {NoStop}%
\bibitem [{\citenamefont {Aleksiej{\=u}nas}\ \emph {et~al.}(2003)\citenamefont
  {Aleksiej{\=u}nas}, \citenamefont {S{\=u}d{\v z}ius}, \citenamefont
  {Malinauskas}, \citenamefont {Vaitkus}, \citenamefont {Jara{\v s}i{\=u}nas},\
  and\ \citenamefont {Sakai}}]{Aleksiejunas_2003}%
  \BibitemOpen
  \bibfield  {author} {\bibinfo {author} {\bibfnamefont {R.}~\bibnamefont
  {Aleksiej{\=u}nas}}, \bibinfo {author} {\bibfnamefont {M.}~\bibnamefont
  {S{\=u}d{\v z}ius}}, \bibinfo {author} {\bibfnamefont {T.}~\bibnamefont
  {Malinauskas}}, \bibinfo {author} {\bibfnamefont {J.}~\bibnamefont
  {Vaitkus}}, \bibinfo {author} {\bibfnamefont {K.}~\bibnamefont {Jara{\v
  s}i{\=u}nas}},\ and\ \bibinfo {author} {\bibfnamefont {S.}~\bibnamefont
  {Sakai}},\ }\bibfield  {title} {\bibinfo {title} {Determination of free
  carrier bipolar diffusion coefficient and surface recombination velocity of
  undoped {{GaN}} epilayers},\ }\href {https://doi.org/10.1063/1.1599036}
  {\bibfield  {journal} {\bibinfo  {journal} {Appl. Phys. Lett.}\ }\textbf
  {\bibinfo {volume} {83}},\ \bibinfo {pages} {1157} (\bibinfo {year}
  {2003})}\BibitemShut {NoStop}%
\bibitem [{\citenamefont {{van Opdorp}}\ \emph {et~al.}(1977)\citenamefont
  {{van Opdorp}}, \citenamefont {Vink},\ and\ \citenamefont
  {Werkhoven}}]{vanOpdorp_1977}%
  \BibitemOpen
  \bibfield  {author} {\bibinfo {author} {\bibfnamefont {C.}~\bibnamefont {{van
  Opdorp}}}, \bibinfo {author} {\bibfnamefont {A.~T.}\ \bibnamefont {Vink}},\
  and\ \bibinfo {author} {\bibfnamefont {C.}~\bibnamefont {Werkhoven}},\
  }\bibfield  {title} {\bibinfo {title} {Minority carrier recombination at
  surfaces, dislocations and microdefects: Evaluation of parameters from near
  band edge luminescence},\ }in\ \href@noop {} {\emph {\bibinfo {booktitle}
  {Inst. {{Phys}}. {{Conf}}. {{Ser}}.}}},\ Vol.\ \bibinfo {volume} {33b}\
  (\bibinfo {year} {1977})\ pp.\ \bibinfo {pages} {317--330}\BibitemShut
  {NoStop}%
\bibitem [{\citenamefont {Lax}(1978)}]{Lax_1978}%
  \BibitemOpen
  \bibfield  {author} {\bibinfo {author} {\bibfnamefont {M.}~\bibnamefont
  {Lax}},\ }\bibfield  {title} {\bibinfo {title} {Junction current and
  luminescence near a dislocation or a surface},\ }\href
  {https://doi.org/10.1063/1.325160} {\bibfield  {journal} {\bibinfo  {journal}
  {J. Appl. Phys.}\ }\textbf {\bibinfo {volume} {49}},\ \bibinfo {pages} {2796}
  (\bibinfo {year} {1978})}\BibitemShut {NoStop}%
\bibitem [{\citenamefont {Donolato}(1998)}]{Donolato_1998}%
  \BibitemOpen
  \bibfield  {author} {\bibinfo {author} {\bibfnamefont {C.}~\bibnamefont
  {Donolato}},\ }\bibfield  {title} {\bibinfo {title} {Modeling the effect of
  dislocations on the minority carrier diffusion length of a semiconductor},\
  }\href {https://doi.org/10.1063/1.368378} {\bibfield  {journal} {\bibinfo
  {journal} {J. Appl. Phys.}\ }\textbf {\bibinfo {volume} {84}},\ \bibinfo
  {pages} {2656} (\bibinfo {year} {1998})}\BibitemShut {NoStop}%
\bibitem [{\citenamefont {Yakimov}(2002)}]{Yakimov_2002}%
  \BibitemOpen
  \bibfield  {author} {\bibinfo {author} {\bibfnamefont {E.~B.}\ \bibnamefont
  {Yakimov}},\ }\bibfield  {title} {\bibinfo {title}
  {Electron-beam-induced-current study of defects in {{GaN}}; experiments and
  simulation},\ }\href {https://doi.org/10.1088/0953-8984/14/48/352} {\bibfield
   {journal} {\bibinfo  {journal} {J. Phys. Condens. Matter}\ }\textbf
  {\bibinfo {volume} {14}},\ \bibinfo {pages} {13069} (\bibinfo {year}
  {2002})}\BibitemShut {NoStop}%
\bibitem [{\citenamefont {Sharma}\ \emph {et~al.}(2000)\citenamefont {Sharma},
  \citenamefont {Thomas}, \citenamefont {Tricker},\ and\ \citenamefont
  {Humphreys}}]{Sharma_2000}%
  \BibitemOpen
  \bibfield  {author} {\bibinfo {author} {\bibfnamefont {N.}~\bibnamefont
  {Sharma}}, \bibinfo {author} {\bibfnamefont {P.}~\bibnamefont {Thomas}},
  \bibinfo {author} {\bibfnamefont {D.}~\bibnamefont {Tricker}},\ and\ \bibinfo
  {author} {\bibfnamefont {C.}~\bibnamefont {Humphreys}},\ }\bibfield  {title}
  {\bibinfo {title} {Chemical mapping and formation of {{V}}-defects in
  {{InGaN}} multiple quantum wells},\ }\href
  {https://doi.org/10.1063/1.1289904} {\bibfield  {journal} {\bibinfo
  {journal} {Appl. Phys. Lett.}\ }\textbf {\bibinfo {volume} {77}},\ \bibinfo
  {pages} {1274} (\bibinfo {year} {2000})}\BibitemShut {NoStop}%
\bibitem [{\citenamefont {Han}\ \emph {et~al.}(2008)\citenamefont {Han},
  \citenamefont {Datta}, \citenamefont {Mahajan}, \citenamefont {Bertram},
  \citenamefont {Lindow}, \citenamefont {Werkhoven},\ and\ \citenamefont
  {Arena}}]{Han_2008}%
  \BibitemOpen
  \bibfield  {author} {\bibinfo {author} {\bibfnamefont {I.}~\bibnamefont
  {Han}}, \bibinfo {author} {\bibfnamefont {R.}~\bibnamefont {Datta}}, \bibinfo
  {author} {\bibfnamefont {S.}~\bibnamefont {Mahajan}}, \bibinfo {author}
  {\bibfnamefont {R.}~\bibnamefont {Bertram}}, \bibinfo {author} {\bibfnamefont
  {E.}~\bibnamefont {Lindow}}, \bibinfo {author} {\bibfnamefont
  {C.}~\bibnamefont {Werkhoven}},\ and\ \bibinfo {author} {\bibfnamefont
  {C.}~\bibnamefont {Arena}},\ }\bibfield  {title} {\bibinfo {title}
  {Characterization of threading dislocations in {{GaN}} using low-temperature
  aqueous {{KOH}} etching and atomic force microscopy},\ }\href
  {https://doi.org/10.1016/j.scriptamat.2008.07.046} {\bibfield  {journal}
  {\bibinfo  {journal} {Scr. Mater.}\ }\textbf {\bibinfo {volume} {59}},\
  \bibinfo {pages} {1171} (\bibinfo {year} {2008})}\BibitemShut {NoStop}%
\bibitem [{\citenamefont {Nakaji}\ \emph {et~al.}(2005)\citenamefont {Nakaji},
  \citenamefont {Grillo}, \citenamefont {Yamamoto},\ and\ \citenamefont
  {Mukai}}]{Nakaji_2005}%
  \BibitemOpen
  \bibfield  {author} {\bibinfo {author} {\bibfnamefont {D.}~\bibnamefont
  {Nakaji}}, \bibinfo {author} {\bibfnamefont {V.}~\bibnamefont {Grillo}},
  \bibinfo {author} {\bibfnamefont {N.}~\bibnamefont {Yamamoto}},\ and\
  \bibinfo {author} {\bibfnamefont {T.}~\bibnamefont {Mukai}},\ }\bibfield
  {title} {\bibinfo {title} {Contrast analysis of dislocation images in
  {{TEM}}\textendash cathodoluminescence technique},\ }\href
  {https://doi.org/10.1093/jmicro/dfi026} {\bibfield  {journal} {\bibinfo
  {journal} {J. Electron Microsc. (Tokyo)}\ }\textbf {\bibinfo {volume} {54}},\
  \bibinfo {pages} {223} (\bibinfo {year} {2005})}\BibitemShut {NoStop}%
\bibitem [{\citenamefont {Pauc}\ \emph {et~al.}(2006)\citenamefont {Pauc},
  \citenamefont {Phillips}, \citenamefont {Aimez},\ and\ \citenamefont
  {Drouin}}]{pauc_2006a}%
  \BibitemOpen
  \bibfield  {author} {\bibinfo {author} {\bibfnamefont {N.}~\bibnamefont
  {Pauc}}, \bibinfo {author} {\bibfnamefont {M.~R.}\ \bibnamefont {Phillips}},
  \bibinfo {author} {\bibfnamefont {V.}~\bibnamefont {Aimez}},\ and\ \bibinfo
  {author} {\bibfnamefont {D.}~\bibnamefont {Drouin}},\ }\bibfield  {title}
  {\bibinfo {title} {Carrier recombination near threading dislocations in
  {{GaN}} epilayers by low voltage cathodoluminescence},\ }\href
  {https://doi.org/10.1063/1.2357881} {\bibfield  {journal} {\bibinfo
  {journal} {Appl. Phys. Lett.}\ }\textbf {\bibinfo {volume} {89}},\ \bibinfo
  {pages} {161905} (\bibinfo {year} {2006})}\BibitemShut {NoStop}%
\bibitem [{\citenamefont {Ino}\ and\ \citenamefont
  {Yamamoto}(2008)}]{Ino_2008}%
  \BibitemOpen
  \bibfield  {author} {\bibinfo {author} {\bibfnamefont {N.}~\bibnamefont
  {Ino}}\ and\ \bibinfo {author} {\bibfnamefont {N.}~\bibnamefont {Yamamoto}},\
  }\bibfield  {title} {\bibinfo {title} {Low temperature diffusion length of
  excitons in gallium nitride measured by cathodoluminescence technique},\
  }\href {https://doi.org/10.1063/1.3040310} {\bibfield  {journal} {\bibinfo
  {journal} {Appl. Phys. Lett.}\ }\textbf {\bibinfo {volume} {93}},\ \bibinfo
  {pages} {232103} (\bibinfo {year} {2008})}\BibitemShut {NoStop}%
\bibitem [{\citenamefont {Yakimov}(2010)}]{Yakimov_2010}%
  \BibitemOpen
  \bibfield  {author} {\bibinfo {author} {\bibfnamefont {E.}~\bibnamefont
  {Yakimov}},\ }\bibfield  {title} {\bibinfo {title} {Comment on ``{{Carrier}}
  recombination near threading dislocations in {{GaN}} epilayers by low voltage
  cathodoluminescence'' [{{Appl}}. {{Phys}}. {{Lett}}. {\textbf{89}}, 161905
  (2006)]},\ }\href {https://doi.org/10.1063/1.3499662} {\bibfield  {journal}
  {\bibinfo  {journal} {Appl. Phys. Lett.}\ }\textbf {\bibinfo {volume} {97}},\
  \bibinfo {pages} {166101} (\bibinfo {year} {2010})}\BibitemShut {NoStop}%
\bibitem [{\citenamefont {Yakimov}(2015)}]{Yakimov_2015}%
  \BibitemOpen
  \bibfield  {author} {\bibinfo {author} {\bibfnamefont {E.~B.}\ \bibnamefont
  {Yakimov}},\ }\bibfield  {title} {\bibinfo {title} {What is the real value of
  diffusion length in {{GaN}}?},\ }\href
  {https://doi.org/10.1016/j.jallcom.2014.11.229} {\bibfield  {journal}
  {\bibinfo  {journal} {J. Alloys Compd.}\ }\textbf {\bibinfo {volume} {627}},\
  \bibinfo {pages} {344} (\bibinfo {year} {2015})}\BibitemShut {NoStop}%
\bibitem [{\citenamefont {Boleininger}\ \emph {et~al.}(2018)\citenamefont
  {Boleininger}, \citenamefont {Swinburne},\ and\ \citenamefont
  {Dudarev}}]{Boleininger_2018}%
  \BibitemOpen
  \bibfield  {author} {\bibinfo {author} {\bibfnamefont {M.}~\bibnamefont
  {Boleininger}}, \bibinfo {author} {\bibfnamefont {T.~D.}\ \bibnamefont
  {Swinburne}},\ and\ \bibinfo {author} {\bibfnamefont {S.~L.}\ \bibnamefont
  {Dudarev}},\ }\bibfield  {title} {\bibinfo {title} {Atomistic-to-continuum
  description of edge dislocation core: Unification of the
  {{Peierls}}-{{Nabarro}} model with linear elasticity},\ }\href
  {https://doi.org/10.1103/PhysRevMaterials.2.083803} {\bibfield  {journal}
  {\bibinfo  {journal} {Phys. Rev. Materials}\ }\textbf {\bibinfo {volume}
  {2}},\ \bibinfo {pages} {083803} (\bibinfo {year} {2018})}\BibitemShut
  {NoStop}%
\bibitem [{Note1()}]{Note1}%
  \BibitemOpen
  \bibinfo {note} {Note that for a nominal temperature of 10~K, the high-energy
  slope of the CL spectra indicates a carrier temperature of 20--30~K. Thus, as
  in Ref.~\protect \citenum {Kaganer_2019}, we have chosen $T=20$~K for the
  calculated profiles.}\BibitemShut {Stop}%
\end{thebibliography}%
\end{document}